\documentclass{article} 

\usepackage{amsmath, amsthm, amssymb, amsfonts}

\usepackage{natbib}
\usepackage[utf8]{inputenc}	
\usepackage[T1]{fontenc}	
\usepackage{xcolor}		
\usepackage[colorlinks = true,
            linkcolor = purple,
            urlcolor  = blue,
            citecolor = cyan,
            anchorcolor = black]{hyperref}	
\usepackage{booktabs} 		
\usepackage{microtype}		
\usepackage{lineno}		
\usepackage{float}			
\usepackage[a4paper,total={17cm,21cm}]{geometry}
\usepackage{multicol}		

\usepackage{tabularx}
\usepackage{subfigure}
\usepackage{libertine}
\usepackage{graphicx}
\usepackage[labelfont=bf,font=small]{caption}

\usepackage{tikz,xcolor,hyperref}

\definecolor{lime}{HTML}{A6CE39}
\DeclareRobustCommand{\orcidicon}{
	\begin{tikzpicture}
	\draw[lime, fill=lime] (0,0) 
	circle [radius=0.16] 
	node[white] {{\fontfamily{qag}\selectfont \tiny ID}};
	\draw[white, fill=white] (-0.0625,0.095) 
	circle [radius=0.007];
	\end{tikzpicture}
	\hspace{-2mm}
}
\foreach \x in {A, ..., Z}{\expandafter\xdef\csname orcid\x\endcsname{\noexpand\href{https://orcid.org/\csname orcidauthor\x\endcsname}
			{\noexpand\orcidicon}}
}

\title{Towards Neural Path Tracing in SRAM}

\author{Mark Pupilli\orcidA{}}
\date{}

\begin{document}

\maketitle
\begin{figure}[H]
  \centering
  \subfigure[Box]{\includegraphics[width=0.19\textwidth]{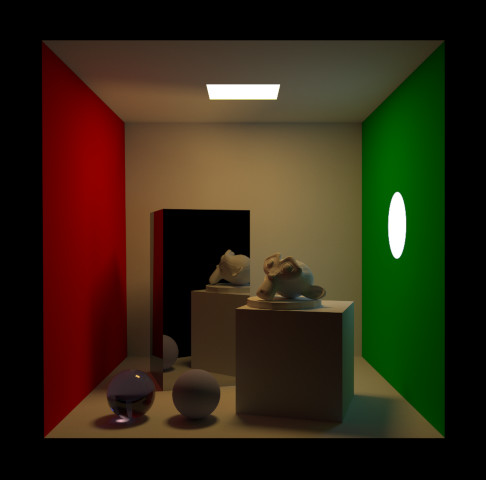}}
  \subfigure[Box + HDR-NIF]{\includegraphics[width=0.19\textwidth]{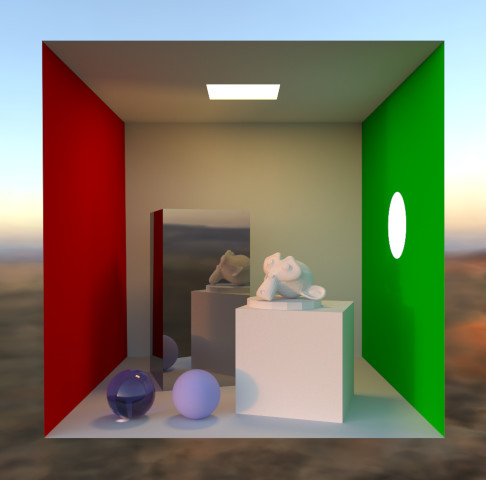}}
  \subfigure[Spheres + HDR-NIF]{\includegraphics[width=0.19\textwidth]{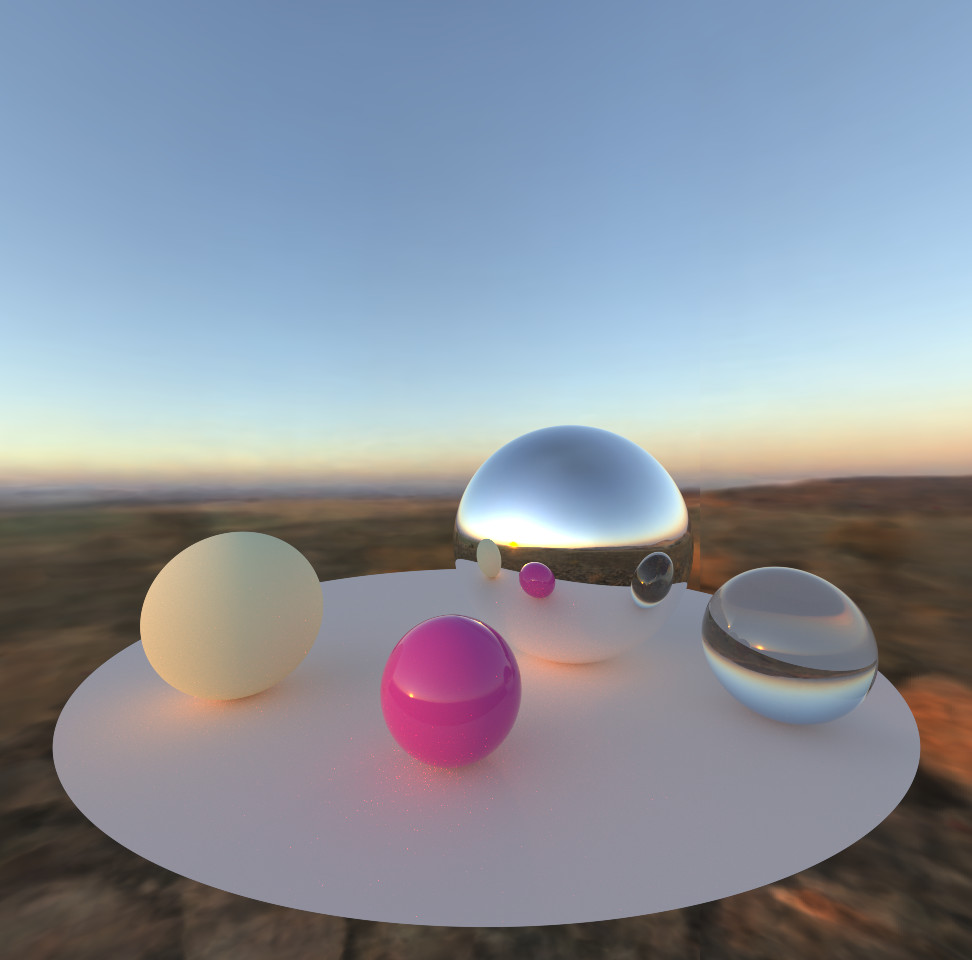}}
  \subfigure[Small BVH + HDR-NIF]{\includegraphics[width=0.1875\textwidth]{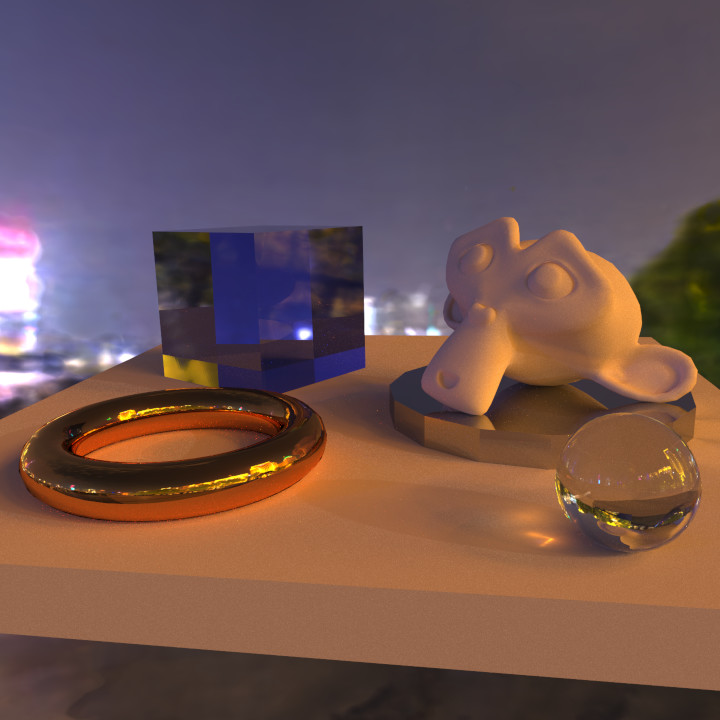}\label{fig:scenes_d}}
  \subfigure[Large BVH]{\includegraphics[width=0.19\textwidth]{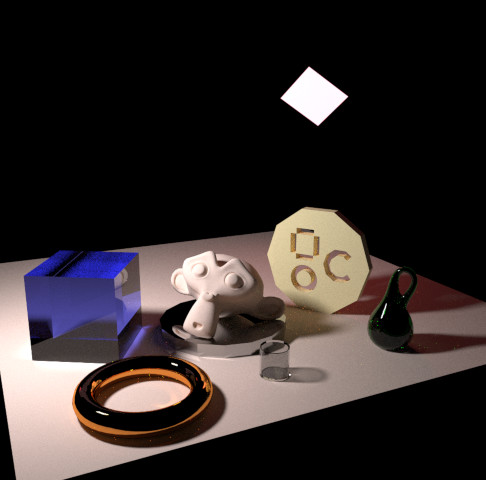}}
  \caption[Second render example.]{Images path traced on a Graphcore Bow-Pod-16. The HDR environment light is compressed into 97 KiB of neural network weights. The neural-network weights, activations, and scene BVH reside entirely in on-chip SRAM.}
  \label{fig:scenes}
\end{figure}

\begin{abstract}
We present an experimental neural path tracer designed to exploit the large on-chip memory of Graphcore intelligence-processing-units (IPUs). This open source renderer demonstrates how to map path tracing to the novel software and hardware architecture and is a useful tool for analysing in-cache neural-rendering scenarios. Such scenarios will be increasingly important if rasterisation is replaced by combinations of ray/path tracing, neural-radiance caching, and AI denoising/up-scaling, for which small neural networks are already routinely employed. A detailed description of the implementation also serves as a self-contained resource for more general software design on IPU.
\end{abstract}

\begin{multicols}{2} 


\section{Introduction}
\label{sec:intro}

Path tracing \cite{Kajiya86} has replaced rasterisation for high-quality offline rendering \cite{Fascione2018, burley2018design, Christensen2018} and real-time rendering seems to be at the start of a similar transition. For example, the ability of GPUs to train and run small neural-networks (NNs) at high throughput and low latency \cite{M_ller_2021_2} allows the bulk of the path tracing workload to be approximated by a neural-radiance-cache \cite{M_ller_2021}. Whilst expensive, NN compute is accelerated by dedicated hardware, amenable to reduced precision arithmetic, and deterministic: all of this is in contrast to the full path tracing algorithm. Neural networks are also utilised for real-time denoising and spatio/temporal super-resolution \cite{yang2023} and on state of the art GPUs up to 88\% of displayed pixels are AI generated as a result \cite{NvidiaDLSS3}. The increased visual fidelity from path tracing and the effectiveness of neural-radiance caches make it a distinct possibility that real-time rendering pipelines become predominantly neural. The key to realising real-time performance for networks in use today is keeping network weights in GPU on-chip memory (SRAM/L1-cache and registers) for as long as possible \cite{M_ller_2021_2, muller2022}. This trend raises the question: do current GPU architectures have the right balance of on-chip memory capacity to off-chip memory bandwidth to adapt to a fundamental shift in the nature of rendering computations?

IPUs were nominally designed for artificial intelligence (AI) and are in no way intended for rendering. That said, each chip is massively parallel like a GPU, but in contrast contains 897MiB SRAM close to the processing cores. Some key architectural differences between IPU and GPU are summarised in Table \ref{table:ipu_vs_gpu}. We are interested in using the large on-chip memory to explore neural rendering configurations that are not currently possible on other hardware. To this end we present an IPU implementation of path tracing combined with a high dynamic range (HDR) neural environment lighting. The network we employ has a similar architecture and size to models used in other neural rendering tasks (see Section \ref{sec:neural-tech}), and those algorithms often additionally involve some form of ray-casting/tracing. For this reason our simple application is a useful tool for reasoning about the performance and viability of in-SRAM neural rendering, regardless of how future hardware adapts.

\begin{table*}[t]
\small
\centering
\begin{tabular}{r|rl}
\textbf{Property} & \textbf{IPU Bow GC200} & \textbf{GPU A100} \\\hline
Independent instruction streams (IIS) & 1472 (tiles) x 6 (threads) = 8832 & 108 (SMs) x 64 (warps) = 6912 \\
SRAM per core & 624 KiB per tile & 192 KiB (up to 164 KiB shared) per SM \\
{\em Contention free} SRAM/registers per IIS & 624/6 = \textbf{104 KiB} & (192 + 256)/64 = \textbf{7 KiB} \\
SIMD vector width per IIS & 2 float & 32 float \\ & 4 half & 64 half \\
Integer compute & Non-vectorized but dual issue & Vectorized but no dual issue\\
Peak TFLOP/sec (half, non-zero) & 349 & 312 \\
\end{tabular}
\caption{Comparison of IPU and GPU (\cite{NvidiaAmpere}) variants with similar silicon area from the same process node (7nm).}
\label{table:ipu_vs_gpu}
\end{table*}

\section{Related Work}

\subsection{Neural Networks in Path Tracing}

Monte Carlo (MC) path tracing \cite{Kajiya86} is a rendering algorithm which accumulates the contributions of all possible light paths in a scene using Monte Carlo integration. It exactly integrates the rendering equation {\em in expectation} and therefore simulates many physical effects producing high fidelity images. Because MC convergence follows an inverse square law it is inevitable that sampling must stop at some point and give way to a denoising algorithm, often implemented using sophisticated AI based denoisers \cite{OpenAIDenoise, zhang2021}. 

Neural networks have also become key to realising real-time path tracing. For example, in \cite{M_ller2020} two neural networks reduce variance of the MC integration: a normalizing-flow network is employed to learn a proposal distribution to enable efficient importance sampling of BSDFs and a second network is used to correct errors introduced by the proposal. In \cite{M_ller_2021} a simpler neural-network is trained online using extended path traces in order that the majority of paths can be stopped after a few bounces, and then fed through a small MLP (en masse) that returns an estimate of the path's remaining radiance. For efficiency, weights are streamed into L1 cache or registers once and then reused repeatedly on the entire batch of rays \cite{M_ller_2021_2}. Depending on batch-size, NN inference throughput for 128 hidden neurons and 2 layers (H128, L2) is reported to be between {\em 280M} and {\em 390M} samples/sec on a 3090 RTX.

\subsection{Other Neural Rendering Techniques}
\label{sec:neural-tech}

Neural radiance fields (NeRFs) \cite{Mildenhall2021} are neural networks trained to approximate continuous functions from $ \mathbb{R}^{5} \mapsto \mathbb{R}^{4} $:
\begin{equation}
    (r, g, b, w) = f(x, y, z, \theta, \phi)
\end{equation}
They map rays in 3D space (point and normalised direction vector in spherical coordinates) into an RGB colour and a weight. This allows a NeRF to encode volumetric data: output colour nominally represents the radiance transmitted along the ray (but in usual practice it is a low dynamic range quantity non-linearly related to luminance). The weight allows the network to represent free space and opacity, but also makes the rendering differentiable allowing it to be part of the training loop. These neural-volume representations have found utility in a number of domains including semantic perception \cite{zhi2021inplace, blomqvist2023neural} and even form the basis for text-to-3D generative AI models \cite{poole2022dreamfusion}.

\subsubsection{Prevalence of the Multi-Layer-Perceptron}
The function NeRFs encode is low dimensional, so small MLP/relu networks ($10^5$ or $10^6$ parameters) give acceptable approximations. In \cite{Mildenhall2021}, for example, the architecture is an MLP with 9 layers and 256 hidden neurons. Deep neural networks exhibit a spectral bias \cite{Rahaman2018} which prevents them from learning high frequency functions (like natural images) without adding engineered features or inductive bias to overcome this. In NeRF the low dimensional input coordinates are embedded into a higher dimensional space using Fourier features \cite{Tancik2020}. SIREN networks \cite{Vincent2020} on the other hand, encourage the network to learn higher frequencies by using sinusoidal activation functions and these networks have the added advantage that they can learn to approximate partial derivatives much more effectively than a similar sized network that uses Fourier features. In CoConet \cite{Bricman2018} they train an MLP network to approximate images but use a hand engineered embedding of pixel coordinates into a higher dimensional input space. A detailed theoretical analysis on alternative input embeddings is given in \cite{Tancik2020}.

Whilst models discussed so far use MLPs, \cite{Minnen2018} employs a convolutional network for image compression. They report the compression artifacts induced by this method are more palatable than JPEG to human observers (smoothed but not blocky) for equivalent bit-rates. Another non-MLP example is the neural texturing sytem of \cite{Thies2019} where a convolutional U-net is used. The network has an encoder/decoder structure hence the rendering system is entirely NN based (16M parameters).

State of the art neural material/texturing systems also utilise MLPs. In \cite{Kuznetsov2021} the problem is again tackled using an MLP that approximates the bi-directional-texture function from $\mathbb{R}^{7} \mapsto \mathbb{R}^{3}$. Most recently \cite{zeltner2023realtime} implements a fully neural material system, combines it with path tracing, and demonstrates improved performance over the traditional shading pipeline: the transition away from rasterisation seems imminent.

Since many techniques in this field use the MLP as a core building block, metrics from our on-chip system should be useful as a baseline for reasoning about the benefits of large on-chip SRAM for a range of neurally based rendering algorithms.

\section{Implementation}
\label{sec:implementation}

We will briefly discuss aspects of the IPU hardware and software architecture relevant to the implementation, more detail is given in Appendix \ref{sec:hw}. The GC200 processor used in this work contains 1472 tiles (cores) and each tile has six, multiple-instruction-multiple-data (MIMD), barrel scheduled, hardware threads called {\em workers}. Each worker issues one instruction packet in round-robin fashion and can dual issue floating point and integer/memory operations. All threads can work in unison to drive each tile's accumulating matrix product (AMP) unit, achieving the peak FLOP rate (Table \ref{table:ipu_vs_gpu}) for matrix multiplies and convolutions. Tiles can synchronise and communicate using a bulk synchronous parallel (BSP) execution scheme that alternates between internal or external exchange of data and local tile computations. The latest variant of the GC200 uses wafer-on-wafer technology and has the designation Bow. Bow IPUs have identical micro-architecture but a 40\% higher clock speed.

\subsection{IPU Path Tracer Design}

The path tracer itself is very simple. Paths are traced through the scene with no light sampling: contributions are only accumulated if paths hit an emissive object by chance or if the path escapes the scene and receives a contribution from the environment light. The environment light field is encoded in a neural network as described in Section \ref{sec:hdr-nif} so the sampling loop alternates between path tracing operations and NN inference. Paths are terminated early by roulette with probability proportional to their radiometric throughput. While the lack of light sampling makes the implementation sample inefficient, it simplifies reasoning about performance.

An entirely on-chip solution needs to store ray payload and results, the scene description, and neural network weights all in SRAM. While 897MiB might seem plenty for this task we need to consider some of the finer details of the architecture. First note the memory is split into 1472 tiles, 624 KiB per core. Cores can not directly access memory of other tiles nor can the cores issue direct accesses to off-chip memory and there is no unified address space. This means that the processor is most efficient when each core is repeatedly accessing data from its local SRAM in parallel and that DRAM access must be batched into a BSP compatible schedule (Appendix \ref{sec:PoplarPrograms}).

\subsubsection{Scene Data}
The first design decision we make is to replicate the scene description on every tile, but spread the neural network weights across the entire chip. The primary reason for this is that routing ray data between tiles containing, for example, different treelets, requires fine grained dynamic exchange patterns. While this is possible in principle using just-in-time (JIT) compiled exchange sequences, this is not exposed in the currently available SDK which only allows for exchanges that are fixed at compile time. Batching ray data for neural network inference has no ordering requirement, so inter-tile exchanges of inputs and results can easily be pre-compiled and in fact the graph compiler can make sophisticated data movement decisions for us (see Section \ref{sec:graph-compile}).

BVH node, triangle, index and vertex buffers are all serialised into a single binary chunk. The IPU programming model is built around variables of a single data type (i.e. tensors) so the idiomatic way of passing the scene data would be using structures of arrays. However, we found this makes the graph description overly verbose and increases the code size needed for inter-tile exchange, so we prefer to serialise all the data into one buffer and pay the cost of de-serialising the data on each tile. De-serialisation takes around 522 worker cycles (3132 total cycles as we only use a single worker). The additional benefit of this scheme is possible experimentation with exchange of treelets \cite{Aila2010} between tiles and/or DRAM in future.

\subsubsection{Ray Data and external DRAM}
\label{sec:rays-and-dram}
We could distribute rays and the frame-buffer across tiles but early experiments with this approach proved limiting in terms of both the resolution of images that can be kept in SRAM and also in terms of ability to store auxiliary payload data with the rays. From an API design point of view it is desirable to associate extra data with each ray, for example normals, hit-point, or other arbitrary output variables. For this reason we choose to stream ray data from external DRAM allowing use of standard ray data structures (84 bytes per ray) instead of a compact representation. DRAM bandwidth in current IPU systems is limited (25.6 GB/sec) but the API allows us to reserve some compute tiles to act as "smart I/O controllers": 32 tiles perform ray loads and stores in parallel with 1440 tiles rendering each ray batch.

The number of rays that each worker thread processes is specified as a graph-compile time constant. This allows us to balance the workload so that DRAM fetches are effectively hidden and large enough ray batches are generated to achieve efficient neural-network inference. (The batch size for each neural network query is {\em rays-per-worker} x 6 x 1440.)

Using DRAM this way allows rendering extremely high resolutions, hiding of DRAM latency, and has another advantage: many standard resolutions are divisible by both 1440 and 6 which ensures all worker threads are utilised (e.g. the 8K format 7680 x 4320 = 3840 x 6 x 1440).

\subsubsection{IPU Ray Tracing Kernels}
\label{sec:kernels}
The ray tracing implementation is plain C++. The only IPU specific code is: 1) use of the vector {\em float2} data type (2 x float32 elements) in the ray-slab intersection test to improve vectorisation; and 2) intrinsics to access the IPU's hardware random number generator inline in the path trace kernel (for example {\em \_\_builtin\_ipu\_urand\_f32()}). We do not yet employ ray-packetisation, which would assist the compiler further with auto-vectorisation. The simplicity is deliberate, partly because we want to share non-obfuscated reference code for IPU graphics programming, and partly because the MIMD nature of the worker threads means there are no execution stream coherence issues we need to manage. As an indication of optimisation potential and generated code quality, the triangle intersection routine (derived from PBRT-v3 \cite{pharr2016physically}) compiles to code with a FLOP arithmetic intensity of 0.23 (67/289) FLOPs per issue with 20/67 of the FLOPs in vectorised instructions. The compiler emits many 32-bit load/stores with only 20/71 using 64-bit. This suggests there is potential to increase FLOP intensity with an IPU specific ray packetisation strategy and optimisations that ensure use of 64/128-bit load/stores. For comparison, the same routine compiled with g++-9.4 targeting x86-64+AVX2 generates 0.30 (67/226) FLOPs/issue, 15 of which are fused multiplies, with no vectorised instructions emitted.

\subsubsection{Bounding Volume Hierarchy}

We found memory optimisations important in building a usable system: next to neural network weights, the bounding volume hierarchy (BVH) is the most significant consumer of tile memory. Our BVH nodes do not differ much from common CPU or GPU implementations with an initial BVH-2 tree built using Embree \cite{wald2014} on the host CPU. This is then compacted into an array of carefully packed, pointer-less nodes, similar to those used in \cite{pharr2016physically}, where the first child is implicitly stored next in the array.

We further reduce memory consumption by storing the extent of the bounding volume in float16 and casting to float32 on-demand at intersection time (the box origins are stored at float32). This reduces node size from 32 bytes to 24. At BVH construction time we use a special software implemented rounding, {\em round-to-nearest-not-lower}, to cast from float32 to float16 which guarantees no intersection can be missed as the bounding box is never smaller than it would have been if stored at float32. Run-time overhead is a negligible cast back from float16 to float32. There are prior examples of compressed BVH nodes \cite{Benthin2018}, with the smallest typically intended for hardware implementation (the feasibility of 1-bit hierarchical encoding is demonstrated in \cite{Keely2014}). Experimenting with sub 16-bit precision seems appealing, however, on IPU casting between integer and floating-point register files can have significant overhead, integer bit manipulation instructions are not vectorised, and the smallest load/stores are 32-bit. Very low precision encoding also increases the total number of intersection tests due to looser bounding boxes. Hence, we prefer the compromise of half precision extents, keeping everything in floating-point. This gives a negligible performance loss due to additional intersections (< 1\% ray throughput on test scenes) yet still gains a 25\% BVH memory saving.

\subsection{Neural Environment Lighting}
\label{sec:hdr-nif}

Our environment lighting model is conceptually like any other neural image field (NIF) approximator. It approximates the low dimensional function: $(r, g, b) = f(u, v)$.

\begin{figure}[H]
  \centering
  \subfigure{\includegraphics[width=0.5\columnwidth]{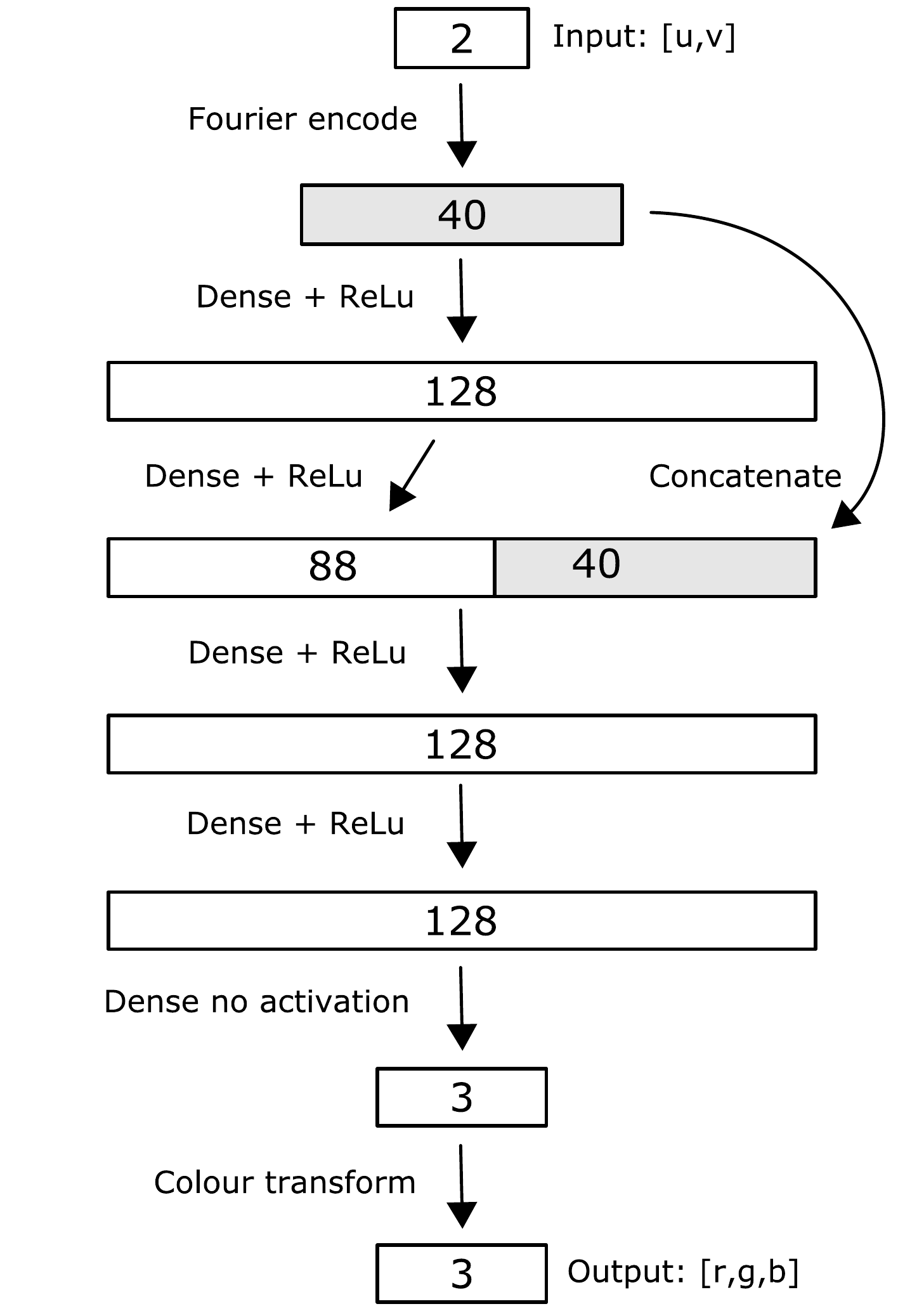}}
  \caption{NIF model schematic showing activation sizes with operations between them.}
  \label{fig:nif-arch}
\end{figure}

The function takes input coordinates $ u, v $, applies positional embeddings as in \cite{Tancik2020} and predicts a colour value using a sequence of dense layers with ReLU activation functions. A schematic for a specific size of the HDR-NIF model is shown in Figure \ref{fig:nif-arch} where hidden size H (128) and number of dense-relu layers L (4) can be varied. The initial Fourier encoded $u,v$ co-ordinates are always concatenated with the last odd layer before the middle of the network, which has a reduced dense layer output size to compensate. The Fourier feature dimension can be varied but we keep it fixed at 40 in all experiments in this paper. The final trainable layer has no activation function allowing the network to regress high dynamic range functions. The network's final layer is a static colour-conversion matrix (defaulting to YUV to RGB). The dynamic range of the RGB training samples is logarithmically compressed as in \cite{LeGendre2019} (during inference we apply the inverse exponential tone-mapping to the final output of the network).

\subsubsection{Training}
Training samples are drawn uniformly on normalised $ u,v $ coordinates and sub-pixel samples are taken from the image using a bi-linear filter. During training we intermittently evaluate the network by feeding $ u, v $ coordinates on a regular grid to reconstruct an image and then compute the PSNR versus the original input image. The model is implemented in Keras and the network is trained on a single GC200 IPU.

We train the network with master weights stored in float16 and enable stochastic rounding \cite{gupta2015deep} which the IPU supports in hardware. This requires a loss scale which we fix at {\em 16384}. We use an ADAM optimiser with {\em learning-rate := 0.001} and internal variables (mean, uncentered-variance) stored at float32. The loss is a Huber loss with {\em delta := 0.001}.

\subsubsection{Approximating HDR Images}

While not the focus of this work some investigation into the behaviour of HDR-NIF convergence was necessary in order to get accurately lit renders and some anecdotal results are included in Section \ref{sec:results}. We discovered that existing training regimes such as \cite{Minnen2018, Bricman2018} do not behave well for high resolution HDR images, so we make two notable adjustments. First we employ a Huber loss, which we found to outperform MSE in terms of accuracy: we hypothesise that encouraging the network to learn high luminance samples more slowly leads to more stable gradients early in training but did not investigate further. We also add a final, fixed, linear colour-space conversion transform to the end of the network: this forces the network to predict samples in a colour-space of our choosing which we found affects accuracy (see Section \ref{colorspace-results}).

\subsubsection{Precision}

The IPU does not have hardware sin/cos instructions. Fourier feature training samples are pre-computed with sin/cos at float32 but during inference, where the Fourier features must be computed on the fly, we use optimised software float16 sin/cos implementations. This does not noticeably affect reconstruction quality but improves performance significantly. In path tracing, where we need float32 $sincos()$, we use an optimised, but generic, software implementation \cite{moshier1992}. This is important because the C++ standard library equivalents use double precision which will be emulated on IPU. Not only is this slow, but the emulation code itself consumes around 10 KiB per tile.

\subsection{Combining Neural-HDRIs and Path Tracing}

Integrating HDR-NIF lighting into the path tracer is conceptually straight forward: to determine the lighting contribution from the environment to a given ray, we just convert its direction in the world coordinate frame to equirectangular $[u,v]$ coordinates and submit these to the NIF which returns the $[r,g,b]$ light contribution to be multiplied with the accumulated ray throughput. For this to be efficient we need to run inference on large batches of ray queries in parallel. The only similar use of NIFs in the literature is \cite{poole2022dreamfusion} where the 3D models are represented using NeRF, but the background is separated, using an neural environment field representation similar to ours. In their case however, the field approximation is low dynamic range.

\subsubsection{Compute Graph Compilation}
\label{sec:graph-compile}

The Poplar/Poplibs graph compilation framework (Appendix \ref{sec:ProgModel}) plans how matrix-multiplies are distributed across cores/tiles automatically. For many model configurations considered here, it is possible that a single layer’s weights fit in one tile's SRAM. In this case Poplibs can decide to replicate weights of individual layers across every tile on-demand, maximising utilisation of the AMP units. It is able to do this because inter-tile bandwidth is high enough that there is a net performance improvement despite the overhead of broadcasting weights before executing each layer. It is interesting to note that Poplibs' matrix-multiply planner is sophisticated enough to make this decision completely automatically: on other platforms this kind of optimisation must be done by hand (as in \cite{M_ller_2021_2}).

When the path trace kernel finishes, the queries it produces are not laid out efficiently for the first matrix multiply in the MLP network. Exchanging information between IPU cores is fast so this is not a problem. However, transferring data between IPUs over IPU link is slower than on-chip exchanges. For this reason when we use multiple IPUs the NIF model is replicated across all the chips which then run data parallel with no communication between them.

\subsubsection{Batch-Serialisation}

The matrix-multiply planner allows us to trade performance versus memory use. If we want to increase performance we can allocate the planner a larger proportion of SRAM for weights and activations. However, when memory use is close to the limit, the optimiser often resorts to serialisation of the output channels of each layer to reduce memory needed for temporary activations. We prefer to manually manage activation memory by batch-serialising the entire network, which we found leads to better inference throughput and memory use in this case.

\section{Results}
\label{sec:results}

All results use the same path tracing parameters: a maximum path length of 10 and roulette termination beginning at depth 3. If the reader wishes to extrapolate performance figures to real-time rendering scenarios they should bare in mind that state of the art real-time path tracing algorithms typically sample one path per-every-other pixel, for a few bounces, then execute one or more neural-networks to complete the task. Here we fully trace paths for every pixel, then execute a neural network, and repeat taking many samples.

\subsection{Test Scenes}

The test scenes evaluated are shown in Figure \ref{fig:scenes}. These scenes are designed to give an indicative spread of results that we can expect for the range of scene sizes and NIF configurations that fit in IPU SRAM. Table \ref{table:scenes} contains some baseline metrics for these scenes. These were measured on an IPU Bow system. In results where we do not specify Bow the performance was measured on classic GC200 chips.

\begin{table}[H]
\scriptsize
\centering
\begin{tabularx}{0.95\columnwidth}{l|rrrrrr}
\textbf{Scene} & Box & Box & Spheres & Small BVH & Large \\
& & + NIF & + NIF & + NIF & BVH\\\hline
\textbf{Giga-paths/sec} & 0.826 & 0.418 & 1.333 & 0.350 & 0.487 \\
\textbf{BVH KiB (per tile)} & 189 & 189 & 0.258 & 265 & 397 \\
\end{tabularx}
\caption{Metrics for test scenes from Figure \ref{fig:scenes} traced on a Bow-Pod-16. NIF model size is 97KiB (H128, L5).}
\label{table:scenes}
\end{table}

\subsection{Precision of Ray Tracing Operations}

In Section \ref{sec:kernels} we mention that some CPU code lowers to AVX fused multiply add/subtract (FMA/S) instructions. The IPU has no infinite intermediate precision FMA/S and we avoided any code that results in fallback to double emulation, so it is prudent to check the accuracy of basic ray tracing operations on IPU. In Table \ref{table:ray-precision} we compare normals and primary hit-points against two CPU implementations: one that uses Embree for ray tracing operations, and one that runs the IPU kernels on CPU. This allows us to distinguish disparities due to algorithmic differences from those due to floating point arithmetic.

We note that differences between the same kernels on IPU/CPU are tiny and consistent, with no difference at all in the primary hits. Against Embree errors are still small but there are notable outliers. The first is the hit-point delta for the box scene, this is because the box scene uses the original Cornell box data with no scaling applied (so dimension range is of the order 100$\times$ larger than the other scenes). The second notable difference is in normals for the small/large BVH scenes compared to box/spheres. In this case the BVH scenes contain many triangles and triangle intersection has the largest algorithmic difference compared to Embree. In effect the larger error is due to outliers where rays hit adjacent (but still valid within machine precision) primitives. In those cases the normal will change dramatically (but the hit-point much less so). Overall, we don't believe these differences should cause any difficulties beyond those normally encountered when ray tracing with single precision.

\begin{table}[H]
\scriptsize
\centering
\begin{tabularx}{\columnwidth}{r|ll|ll}
\textbf{Scene} & \textbf{MSE (vs Embree)} & & \textbf{MSE (vs CPU)} \\\hline
 & Normal & Hit & Normal & Hit \\
Box & $1.1 \times 10^{-13}$ & $7.6 \times 10^{-9}$ & $2.6 \times 10^{-16}$ & $0$ \\
Spheres & $2.1 \times 10^{-14}$ & $2.5 \times 10^{-14}$ & $1.6 \times 10^{-16}$ & $0$ \\
Small BVH & $4.5 \times 10^{-7}$ & $2.3 \times 10^{-14}$ & $1.8 \times 10^{-16}$ & $0$ \\
Large BVH & $1.2 \times 10^{-7}$ & $7.1 \times 10^{-14}$ & $2.2 \times 10^{-16}$ & $0$ \\
\end{tabularx}
\caption{Precision test results showing the worst (across the $x,y,z$ components) mean squared error (MSE) for normals and primary hit-points.}
\label{table:ray-precision}
\end{table}

\subsection{Colourspaces}
\label{colorspace-results}

We found that images with significant dynamic range require careful selection of the colour-space in which the NN learns its representation. This problem is exacerbated with smaller batch-sizes ($O(10^3)$). In extreme examples, such as where midday sun is the only light source, training can be very unstable if the network learns in the RGB colour-space. In this case all the energy in the scene comes from $<<$ 1\% of the HDRI's pixels: in effect the most important training samples are outliers.

\begin{figure*}[t]
  \centering
  \subfigure[Network learned RGB]{\label{fig:colorspace-a}\includegraphics[width=0.22\textwidth]{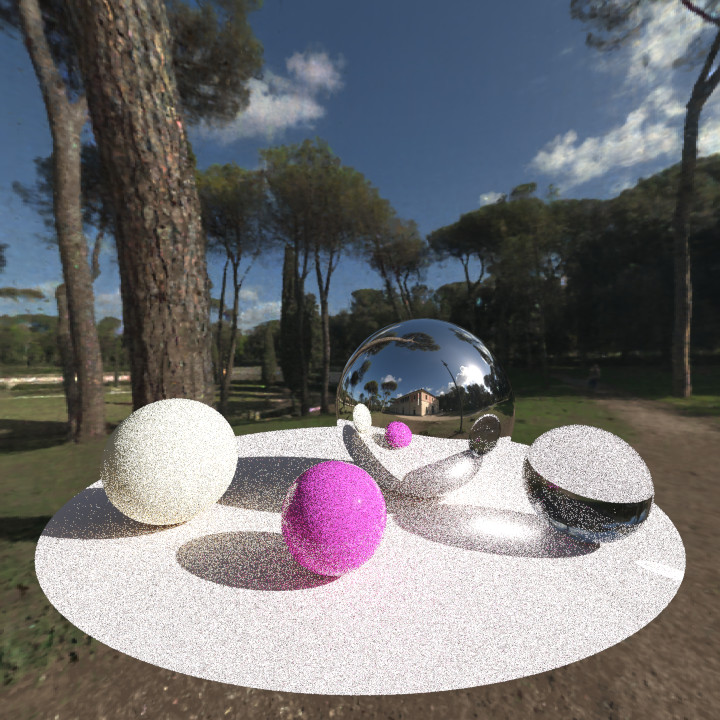}}
  \subfigure[Network learned YCoCg]{\label{fig:colorspace-b}\includegraphics[width=0.22\textwidth]{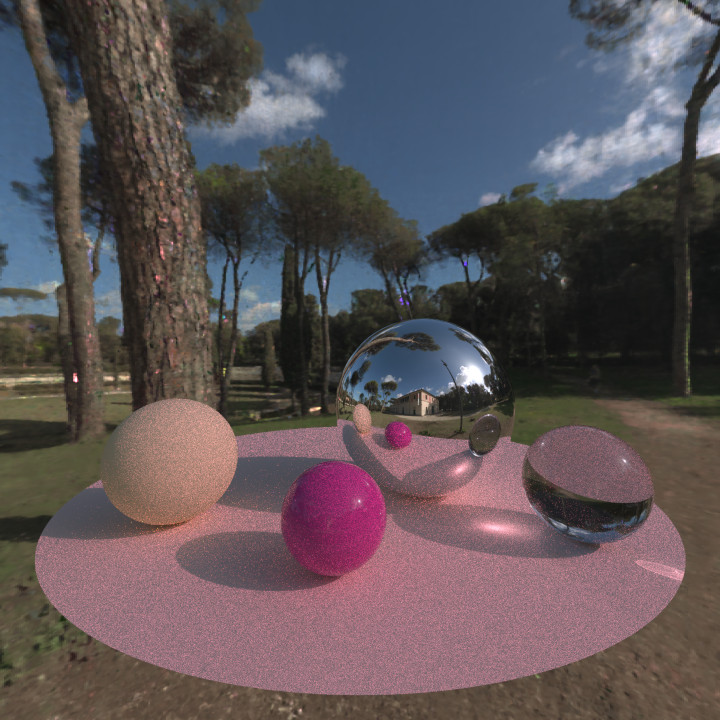}}
  \subfigure[Network learned YUV]{\label{fig:colorspace-c}\includegraphics[width=0.22\textwidth]{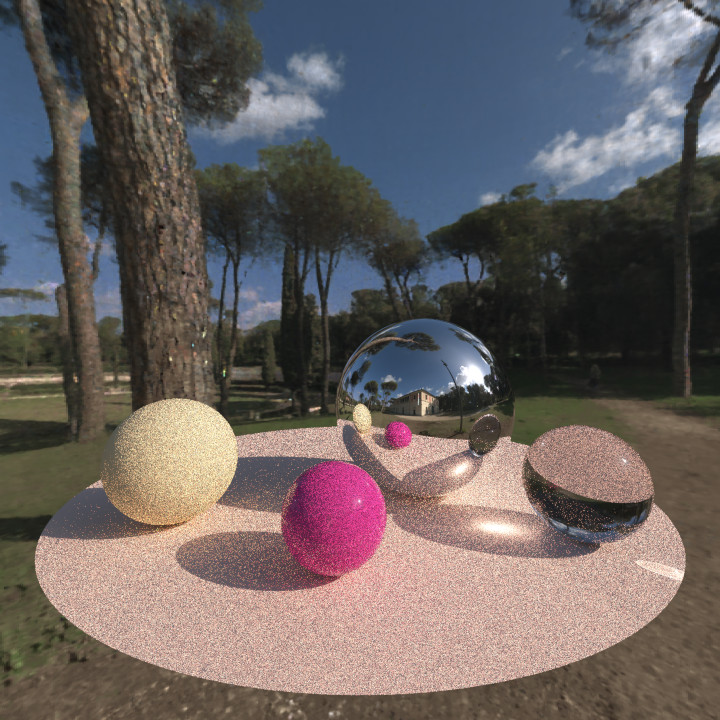}}
  \caption{Images path traced using HDR-NIFs trained in different colour-spaces. (Network that learns a YUV representation gives light with the correct hue).} \label{fig:colorspace}
\end{figure*}

Results of forcing the network to learn in different colour-spaces with a sun-lit HDRI are shown in Figure \ref{fig:colorspace}. Table \ref{table:colorspace} shows the peak-signal-to-noise-ratio (PSNR) of the reconstructed NIF versus the orignal HDRI. Note the background where the NIF is directly hit by primary rays: it contains few perceptible differences between the three images because the errors in the few sun samples are only noticeable when used for path tracing. Only YUV gives the correct result in this case, despite YCoCg having higher PSNR in luminance channel and RGB space. The YCoCg colour matrix completely de-correlates gradients flowing back from the loss layer whilst in contrast, the RGB colour-space is known to have highly correlated components. We hypothesise that having some correlation between gradients is beneficial, but the network is not able to separate luminance and chrominance information from RGB samples alone. We believe these aspects warrant further investigation.

\begin{table}[H]
\small
\centering
\begin{tabular}{r|lll}
\textbf{Color-space} & \textbf{PSNR} & \textbf{PSNR} & \textbf{PSNR} \\
& \textbf{RGB} & \textbf{Luminance} & \textbf{Chrominance} \\\hline
RGB & 32.4 & 37.8 & 28.1 \\
YCoCg & \textbf{57.8} & \textbf{61.5} & 43.9 \\
YUV & 57.7 & 61.4 & \textbf{45.7} \\
\end{tabular}
\caption{PSNRs for Borghese Gardens HDR-NIF approximation using different colour spaces. Highest PSNRs in bold.}
\label{table:colorspace}
\end{table}

\subsection{NIF Accuracy}

\begin{figure*}[t]
  \centering
  \subfigure[H64 L2, 311.8M paths/sec.]{\label{fig:nrfa}\includegraphics[width=0.22\textwidth]{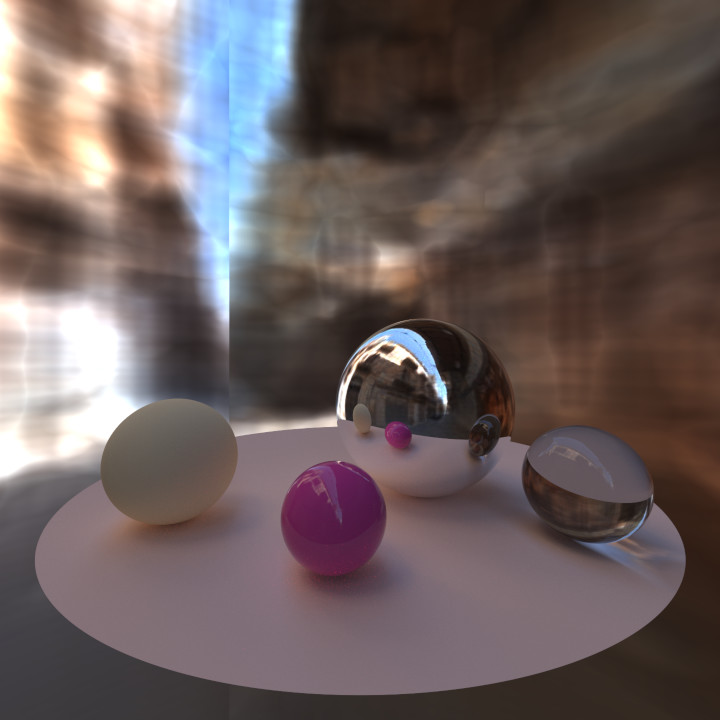}}
  \subfigure[H256 L4, 149.1M paths/sec.]{\label{fig:nrfb}\includegraphics[width=0.22\textwidth]{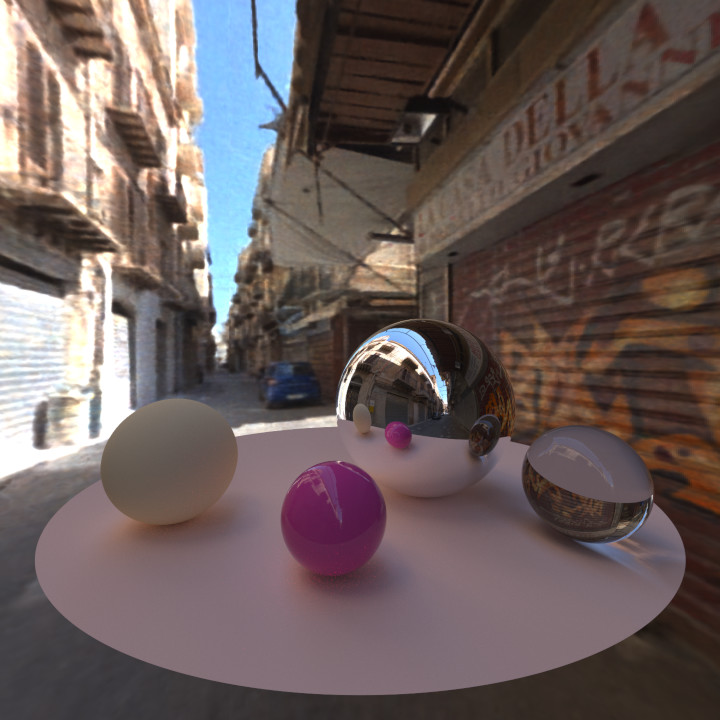}}
  \subfigure[H1024 L8, 9.3M paths/sec.]{\label{fig:nrfc}\includegraphics[width=0.22\textwidth]{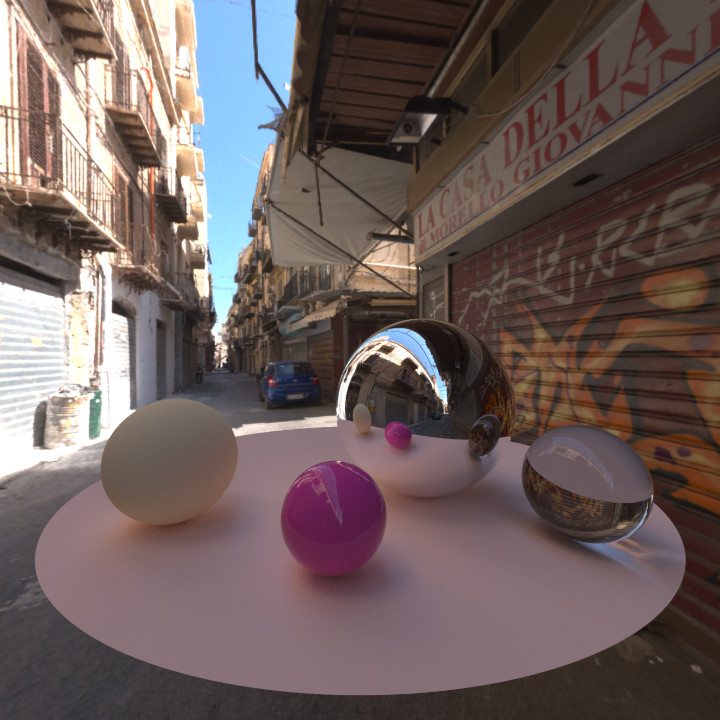}}
  \subfigure[Reference (original 4K HDRI).]{\label{fig:nrfd}\includegraphics[width=0.22\textwidth]{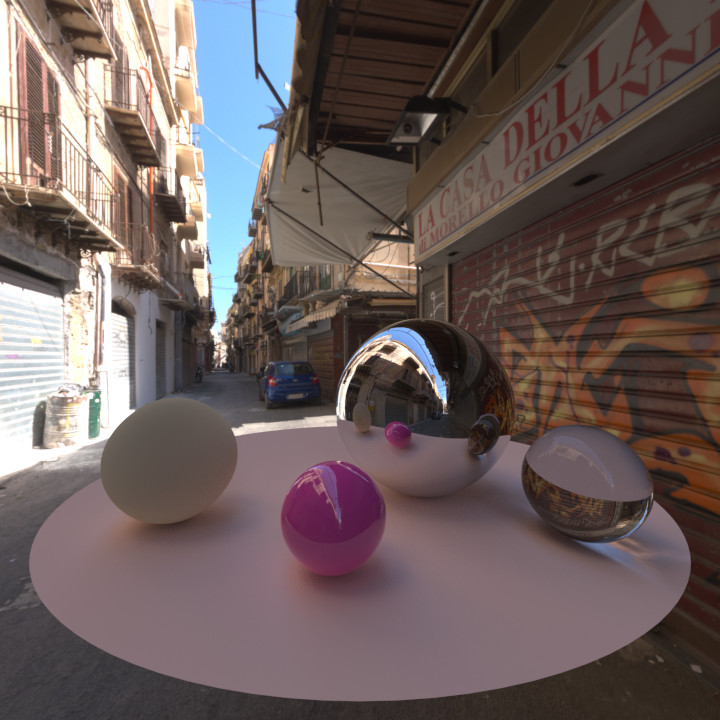}}
  \caption{Comparison of neural path tracing results with different HDR-NIF model sizes for the {\em urban alley} HDRI.} \label{fig:nrf-vs-blender}
\end{figure*}

In Figure \ref{fig:nrf-vs-blender} we can see a comparison of path tracing results for the {\em spheres} scene with HDR-NIFs of increasing model size and a reference image rendered in Blender \cite{blender2018}. This simple scene is used so that we can vary the NIF model size through a large range without exhausting SRAM. In Figure \ref{fig:nrfa} even though the background is a very poor approximation of the original HDRI the shadows and caustics look comparable to the higher quality NIFs and original reference in \ref{fig:nrfd}. This explains why tiny neural radiance caches can be so effective: because in that use case the neural-field approximation is only used to complete paths and never directly visualised.

We list the PSNR values for the {\em urban alley} HDRI in Table \ref{table:model_vs_psnr} because it is an outlier in terms of other PSNR results. Figure \ref{fig:rate-psnr-plots} shows model-sizes versus PSNRs and path rate (paths/second) for a sweep over 7 different HDRI images: {\em urban alley} can be seen as the left most vertical grouping of scatter points with significantly lower PSNR than the rest. This may be because it has high texture and colour variation as well as large dynamic range. In this case we note that the chromatic PSNR is a poor predictor of reconstruction quality which is at odds with Table \ref{table:colorspace}.

\begin{table}[H]
\small
\centering
\begin{tabular}{r|lll}
\textbf{Model Config} & \textbf{PSNR} & \textbf{PSNR} & \textbf{PSNR} \\
& \textbf{RGB} & \textbf{Luminance} & \textbf{Chrominance} \\\hline
H64 L2 & 29.0 & 31.1 & \textbf{14.5} \\
H256 L4 & 29.3 & 35.6 & 14.1 \\
H1024 L8 & \textbf{29.4} & \textbf{37.3} & 14.0 \\
\end{tabular}
\caption{PSNRs for Urban Alley HDR-NIF approximations. Highest PSNRs in bold.}
\label{table:model_vs_psnr}
\end{table}

We are aware of perceptual HDR metrics, such as \cite{Mantiuk2011}, which should be preferred over PSNR in some circumstances. However, here we are jointly interested in the perceptual reconstruction of the directly visible background, and indirect effects that affect validity of the lighting simulation. It is not immediately clear how perceptual metrics help us in this case. For the remainder of the paper we continue to report three variations of PSNR but recognise their limited utility.

\begin{figure}[H]
  \centering
  \subfigure{\includegraphics[width=0.9\columnwidth]{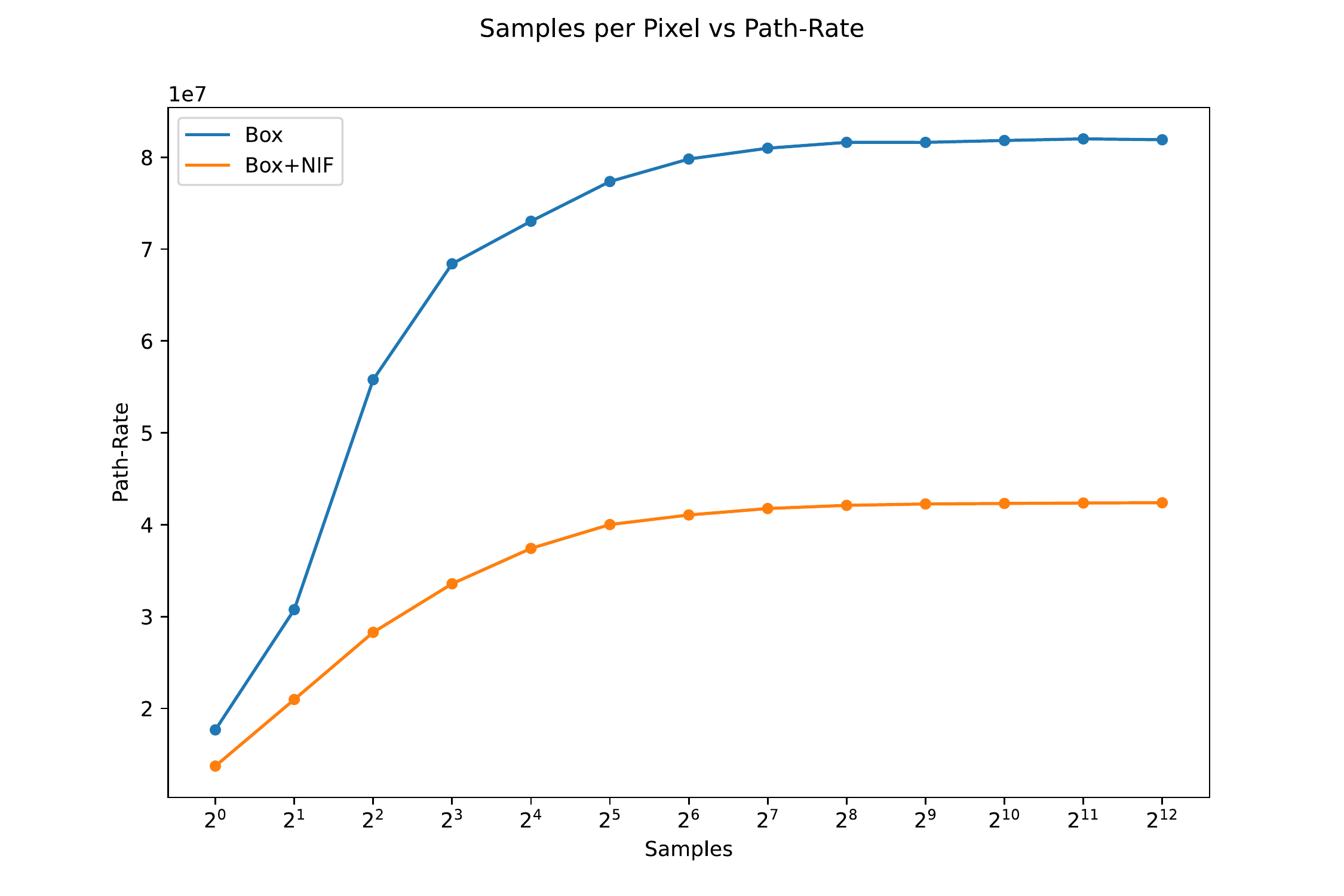}}
  \caption{Effect of sample count on path rate (paths/sec), with and without an HDR-NIF.}
  \label{fig:roofline}
\end{figure}

\subsection{Performance}

Figure \ref{fig:roofline} shows a roofline plot of sample count versus path rate when rendering 720 $\times$ 720 images of the box scene on a single IPU GC200 chip from a Pod-4 (classic) system. Even with relatively low bandwidth the transfers can be completely overlapped with compute when taking more than 256 samples per pixel. The DRAM ray transfer bandwidth is ultimately limited by transfer size, which in turn is limited by the amount of ray-data that fits on 32 I/O tiles (see Table \ref{table:DRAM-BW}). Note that ray batch data is double buffered on the I/O tiles to allow saving one batch while reading the next efficiently in a pipeline.

\begin{table}[H]
\footnotesize
\centering
\begin{tabular}{r|lll}
\textbf{Image} & \textbf{Save+Load Time} & \textbf{Bytes per} & \textbf{Bandwidth)}\\
\textbf{Resolution} & \textbf{(milli-seconds)} & \textbf{Ray-Batch} & \textbf{(GiB/sec)}\\\hline
360$\times$360 & 3.9 & 5443200 & 1.29 \\
720$\times$720 & 4.0 & 5443200 & 1.27 \\
1440$\times$1440 & 3.3 & 5806080 & 1.64 \\
2880$\times$2880 & 3.3 & 5806080 & 1.64 \\
5760$\times$5760 & 3.3 & 7257600 & 1.65 \\
\end{tabular}
\caption{Achieved bi-directional DRAM ray transfer bandwidth when using 8 {\em rays-per-worker} thread.}
\label{table:DRAM-BW}
\end{table}

\begin{figure*}
  \subfigure[]{\label{fig:rpwa}\includegraphics[width=0.49\textwidth]{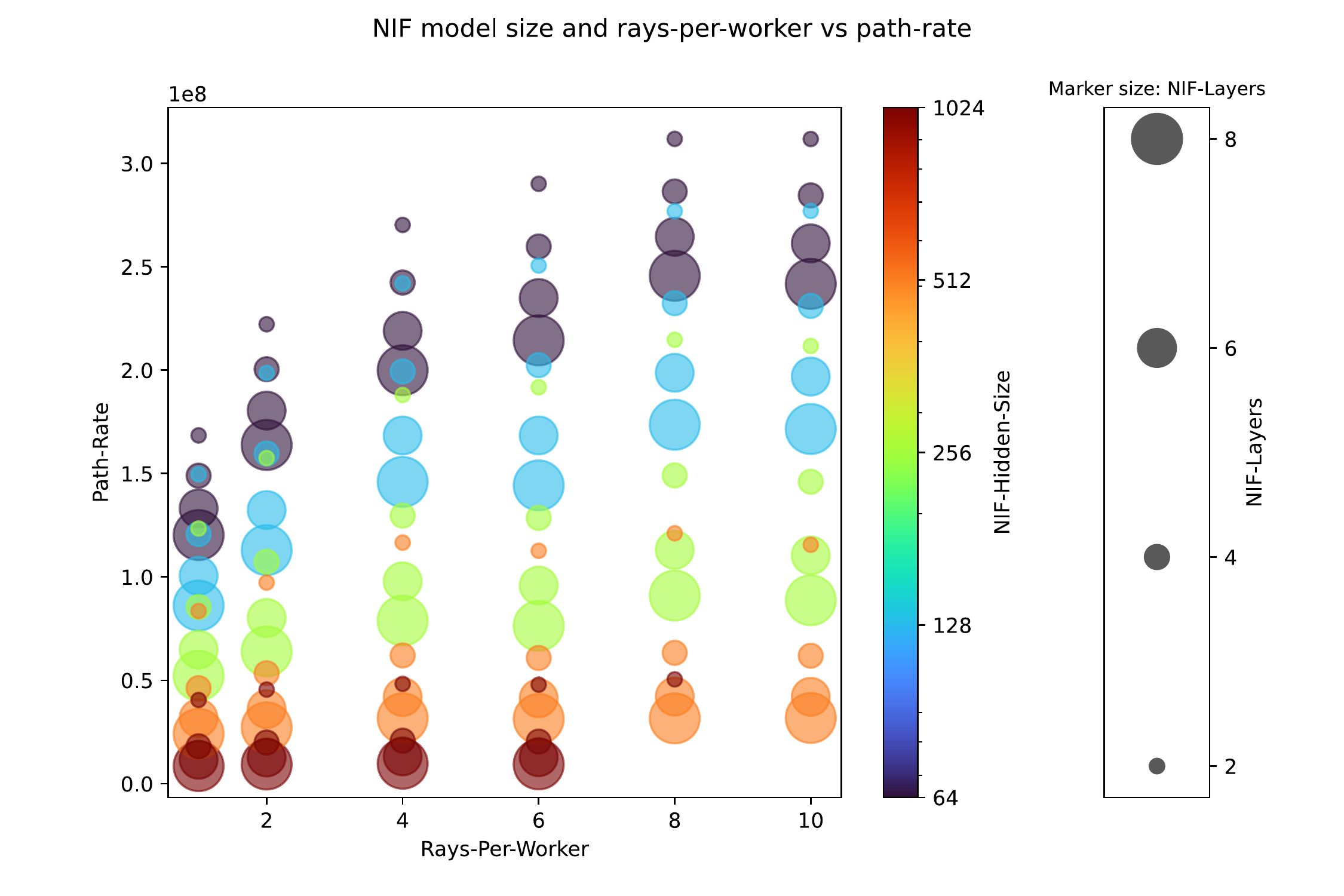}}
  \subfigure[]{\label{fig:rpwb}\includegraphics[width=0.49\textwidth]{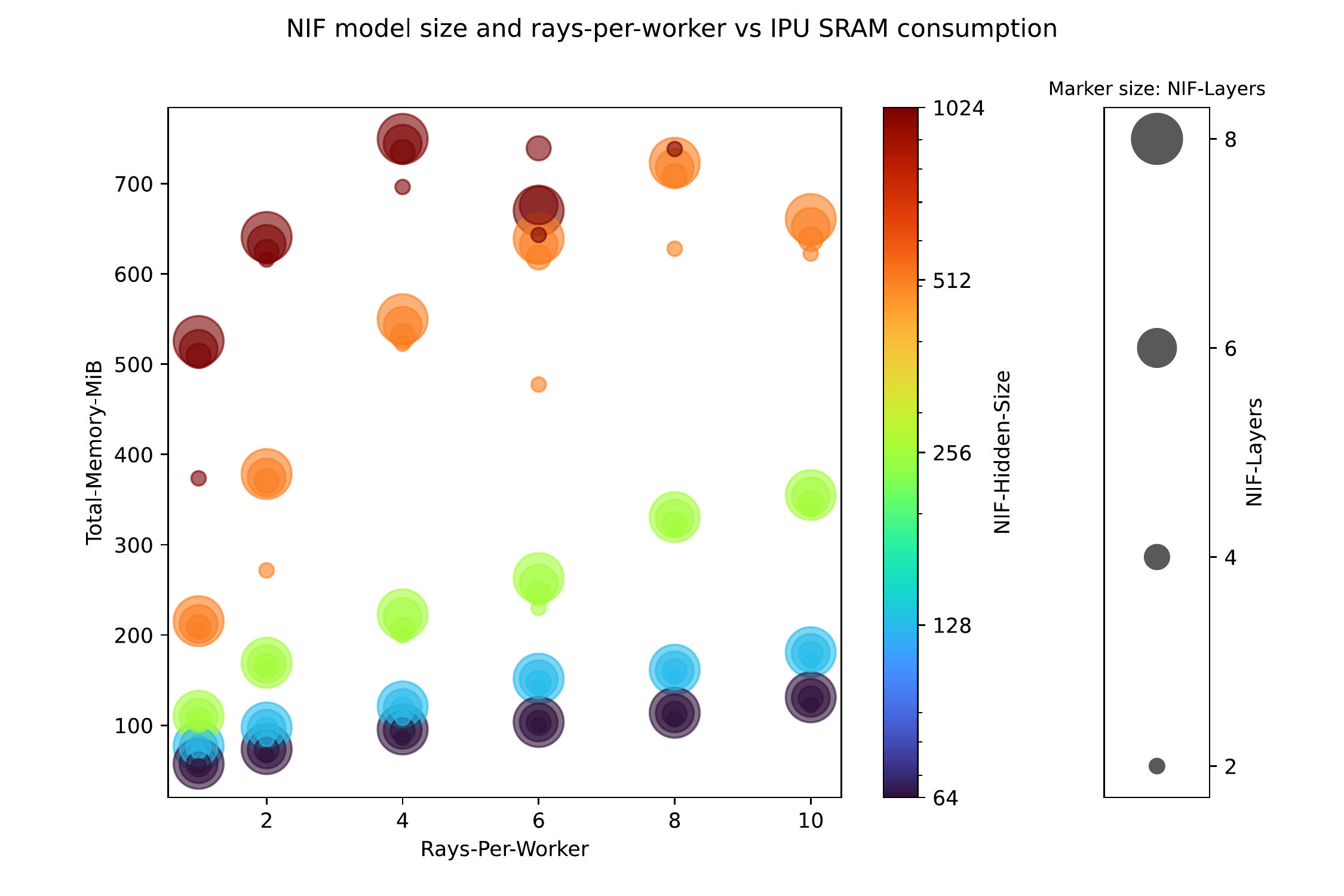}}
  \caption{4D scatter plots showing the effect of NIF model configuration (layers and hidden-size) and rays-per-worker thread on path throughput and on-chip memory consumption.} \label{fig:rpw-plots}
\end{figure*}

Bytes per ray batch in Table \ref{table:DRAM-BW} does not exhaust I/O tile memory so in principle, we could increase the DRAM bandwidth slightly by increasing the batch size, which amounts to increasing {\em rays-per-worker} (Section \ref{sec:rays-and-dram}). However, in Figure \ref{fig:rpwa} we can see that overall path trace rate peaks at 8 {\em rays-per-worker} (and this seems consistent across many NIF configurations). So unless very small numbers of samples are required, there is no performance benefit from increasing rays-per-worker beyond 8, despite this limiting the DRAM bandwidth utilisation. Furthermore, increasing {\em rays-per-worker} to 10 exhausts SRAM for larger NIF models (Figure \ref{fig:rpwb}) so there is limited room for manoeuvre and we generally stick to 6 or 8 rays-per-worker.

\subsubsection{Scaling with Clock Speed}
Bow IPU systems have identical micro-architecture to IPU-Classic systems but a 40\% higher clock speed. If we take enough samples to maintain overlapped DRAM transfers then we would expect all on-chip operations to scale perfectly with clock speed. We measure a single test case (Figure \ref{fig:scenes_d}) on a Bow system to check this. Results are given in Table \ref{table:clock_scaling} where samples per clock are almost identical between the two systems, demonstrating the expected scaling.

\begin{table}[H]
\footnotesize
\centering
\begin{tabular}{r|lll}
\textbf{IPU} & \textbf{Sample-rate} & \textbf{Clock} & \textbf{Sample-rate}\\
\textbf{System} & \textbf{(paths/sec)} & \textbf{(GHz)} & \textbf{per GHz}\\\hline
Pod-4 Classic & 156.5 & 1.33 & 117.7 \\
Bow Pod-4 & 216.4 & 1.85 & 117.0 \\
\end{tabular}
\caption{Scaling of neural path trace rate with clock-speed.}
\label{table:clock_scaling}
\end{table}

\subsection{Model-size/PSNR versus Path Rate}

\begin{figure*}
  \subfigure[]{\label{fig:ratea}\includegraphics[width=0.49\textwidth]{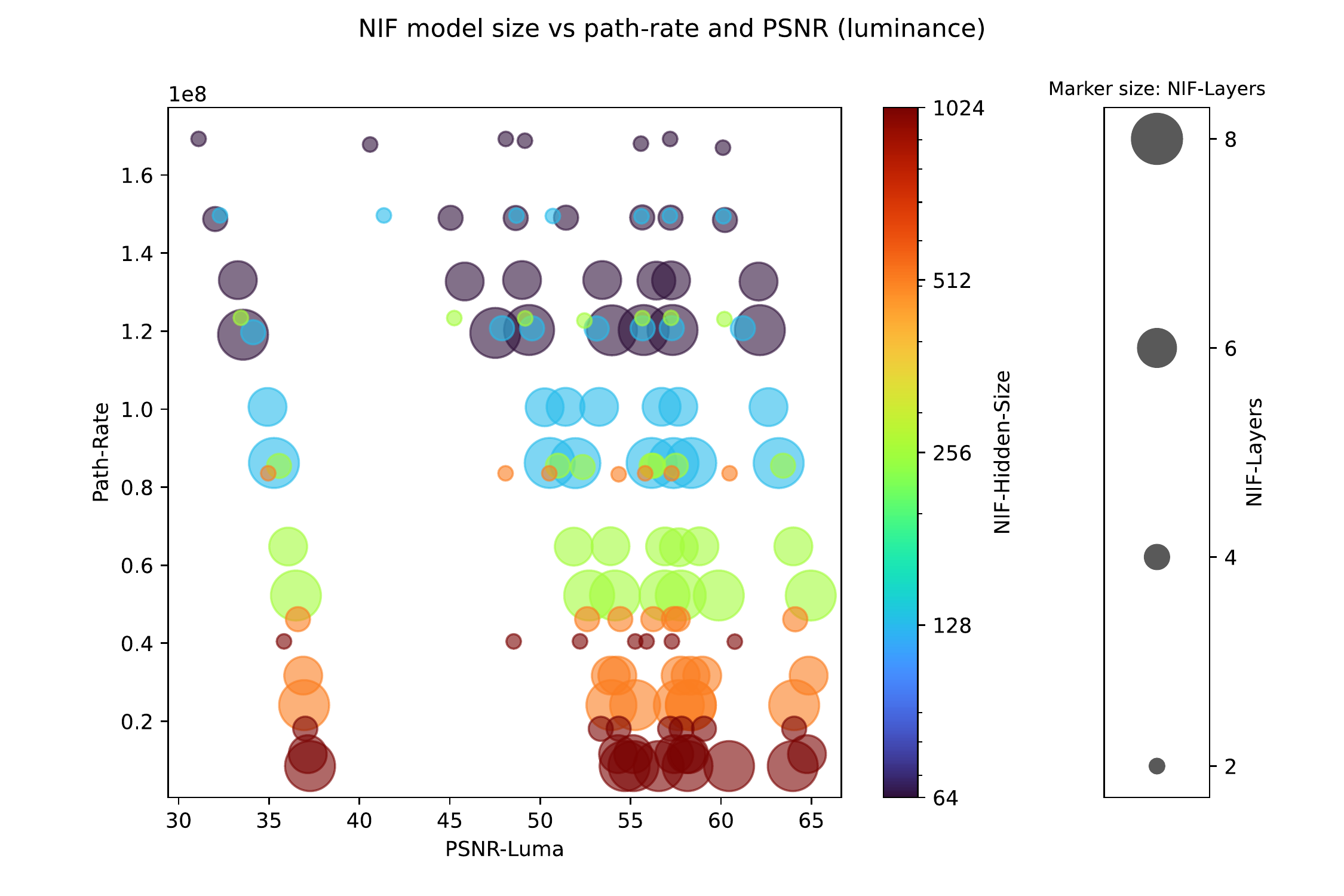}}
  \subfigure[]{\label{fig:rateb}\includegraphics[width=0.49\textwidth]{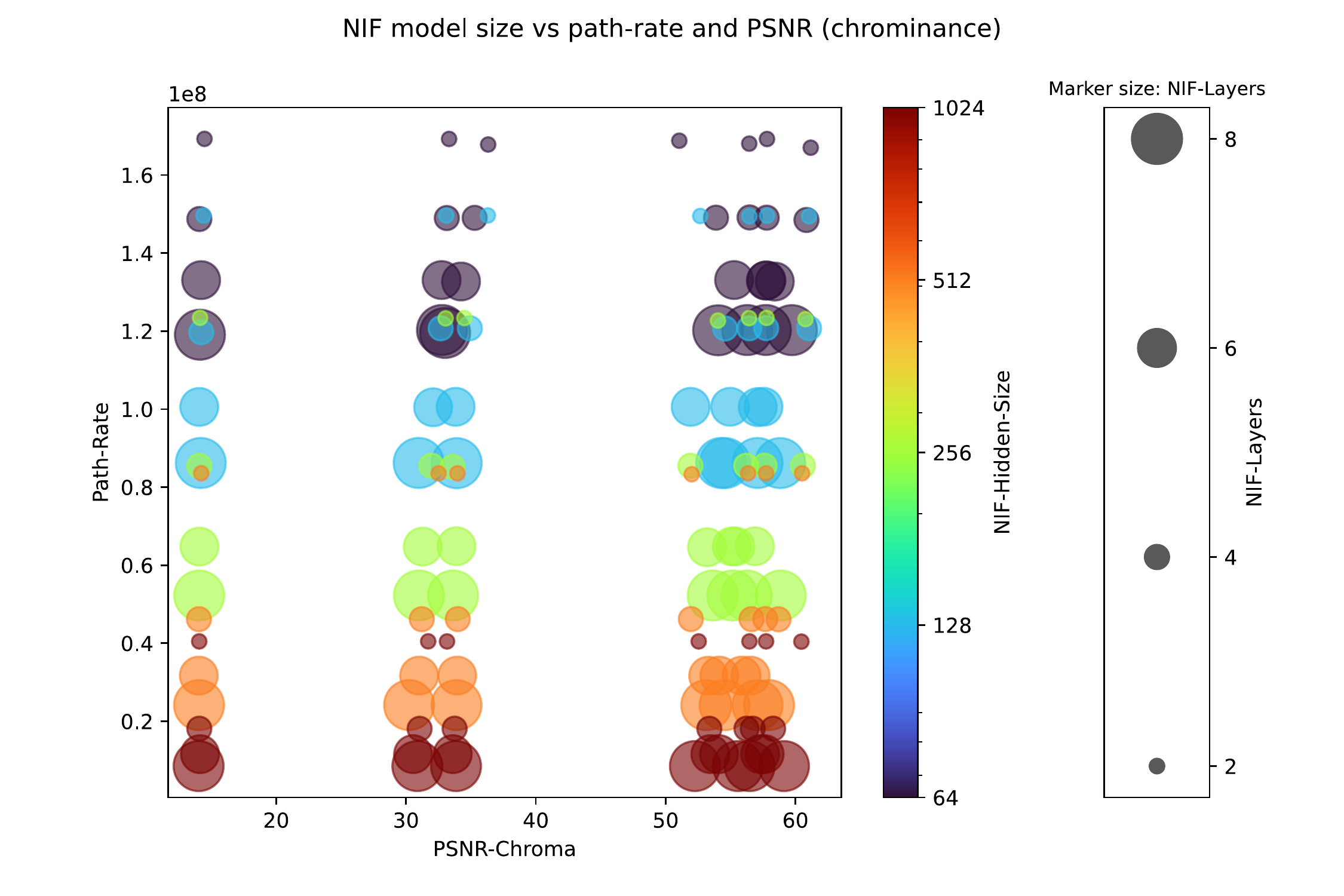}}
  \subfigure[]{\label{fig:ratec}\includegraphics[width=0.49\textwidth]{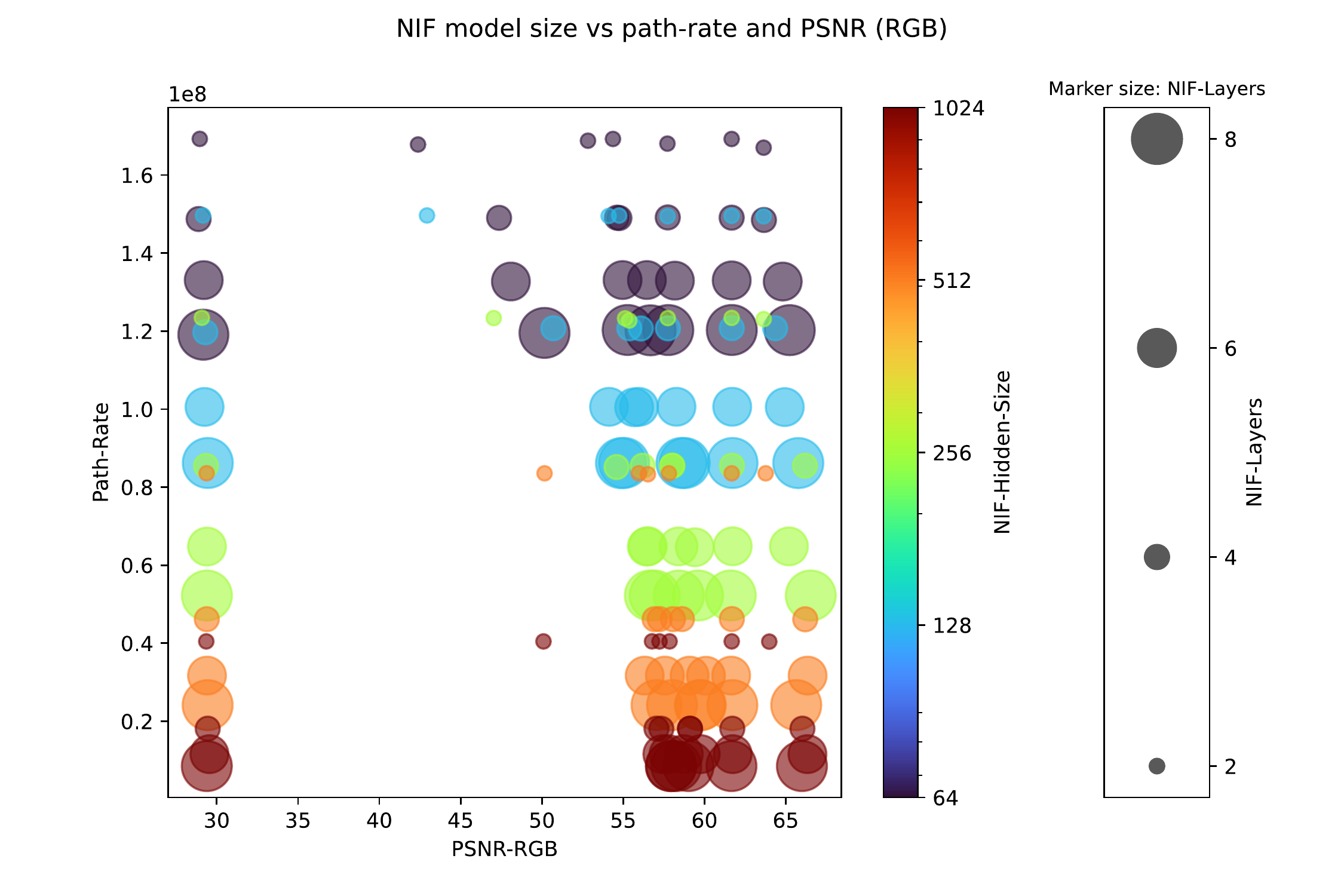}}
  \caption{4D scatter plots showing how NIF model size influences path rate (paths/sec) and PSNR (in luminance \ref{fig:ratea}, chrominance \ref{fig:rateb}, and RGB \ref{fig:ratec}).}
  \label{fig:rate-psnr-plots}
\end{figure*}

Figure \ref{fig:rate-psnr-plots} shows how NIF model configuration affects path sampling rate (paths/sec) and PSNR (dB). Sample rates are again for a single GC200. As well as sweeping across NIF size we also sweep across 7 HDR-NIFs, trained on the HDRIs listed in Appendix \ref{sec:hdri-assets}.

Unfortunately, chromatic PSNR (Figure \ref{fig:rateb}) appears insensitive to model size, despite it having a large effect on correctness of the lighting, but we have seen that lighting can be affected by outlying samples that will not be captured by PSNR metrics (Figure \ref{fig:colorspace}). Luminance (Figure \ref{fig:ratea}) is more sensitive: only models with either hidden size above 128, or layer counts above 4, (or both), reliably result in PSNRs above 60. These are interesting thresholds because they are precisely the points at which current GPUs can not operate in the regime that loads weights into SRAM/registers exactly once \cite{M_ller_2021_2}. Of course, things are not so clear cut, as we have seen in Figure \ref{fig:nrfa}, path tracing effects can have high fidelity even when the background reconstruction quality is unacceptable.

\section {Conclusions and Future Work}

We have presented results that demonstrate the potential and utility of Graphcore processors for exploring on-chip neural rendering techniques. The current system is effective when all data, other than ray streams, fits in SRAM. The capability to render, even small scenes, entirely from SRAM is a capability unique to IPUs and we believe it is worth exploring further.

Functionally, the requirement for all BVH data to fit on one tile is currently the most significant limitation. Algorithms for distributed rendering between thousands of CPUs, such as \cite{Fouladi2022}, may well be applicable at a smaller scale within a single IPU processor, and we would like to explore similar algorithms to exchange both treelets and rays between tiles, taking advantage of the high on-chip communication bandwidth to render larger scenes. Using the currently idle compute capability of the I/O tiles is also an interesting possibility, for example to decompress treelets, or to sort and schedule rays as they are loaded in parallel with the heavier compute.

In terms of performance there is scope for optimisation. For example, tracing ray packets of size two and storing more than one triangle primitive per BVH leaf-node would drastically increase the scope for vectorisation and reduce BVH memory consumption. Combined with more judicious use of larger load/store instructions it should be possible to increase FLOP intensity. Newer IPU hardware supports float8 and storing the HDR-NIF weights in this format would halve the memory required for weights and increase performance of the system's neural component. The BVH representation could also be compressed further using quarter-precision.

Finally, a more sophisticated material and shading system including textures and importance sampling would be a desirable addition. In current systems the lack of fine grained, random external DRAM reads for shade on hit texturing suggests a {\em shade before hit} approach \cite{Fascione2018}, or alternatively, fully neural material and geometry systems \cite{park2019, Kuznetsov2021, zeltner2023realtime} may be more appropriate given that IPUs are predominantly AI processors.

Hopefully the work presented here encourages others to experiment with IPU hardware in this domain with confidence in the basic building blocks we have provided: all source code is made available here \cite{ipuraylib2023}. It is probably not yet clear which types and sizes of neural networks rendering hardware should be optimised for, but we believe the IPU is a useful tool in evaluating the range of possibilities.

\section {Acknowledgements}
We would like to thank Deniz Beker, Yani Donchev, Andrew Fitzgibbon, and Carlo Luschi for their input and feedback.

{\footnotesize
\bibliography{main}}

\begin{thebibliography}{43}
\providecommand{\natexlab}[1]{#1}
\providecommand{\url}[1]{\texttt{#1}}
\expandafter\ifx\csname urlstyle\endcsname\relax
  \providecommand{\doi}[1]{doi: #1}\else
  \providecommand{\doi}{doi: \begingroup \urlstyle{rm}\Url}\fi

\bibitem[Kajiya(1986)]{Kajiya86}
James~T. Kajiya.
\newblock The rendering equation.
\newblock \emph{SIGGRAPH Comput. Graph.}, 20\penalty0 (4):\penalty0 143–150,
  August 1986.
\newblock ISSN 0097-8930.
\newblock \doi{10.1145/15886.15902}.
\newblock URL \url{https://doi.org/10.1145/15886.15902}.

\bibitem[Fascione et~al.(2018)Fascione, Hanika, Leone, Droske, Schwarzhaupt,
  Davidovi\v{c}, Weidlich, and Meng]{Fascione2018}
Luca Fascione, Johannes Hanika, Mark Leone, Marc Droske, Jorge Schwarzhaupt,
  Tom\'{a}\v{s} Davidovi\v{c}, Andrea Weidlich, and Johannes Meng.
\newblock Manuka: A batch-shading architecture for spectral path tracing in
  movie production.
\newblock \emph{ACM Trans. Graph.}, 37\penalty0 (3), aug 2018.
\newblock ISSN 0730-0301.
\newblock \doi{10.1145/3182161}.
\newblock URL \url{https://doi.org/10.1145/3182161}.

\bibitem[Burley et~al.(2018)Burley, Adler, Chiang, Driskill, Habel, Kelly,
  Kutz, Li, and Teece]{burley2018design}
Brent Burley, David Adler, Matt Jen-Yuan Chiang, Hank Driskill, Ralf Habel,
  Patrick Kelly, Peter Kutz, Yining~Karl Li, and Daniel Teece.
\newblock The design and evolution of disney’s hyperion renderer.
\newblock \emph{ACM Transactions on Graphics (TOG)}, 37\penalty0 (3):\penalty0
  1--22, 2018.

\bibitem[Christensen et~al.(2018)Christensen, Fong, Shade, Wooten, Schubert,
  Kensler, Friedman, Kilpatrick, Ramshaw, Bannister, Rayner, Brouillat, and
  Liani]{Christensen2018}
Per Christensen, Julian Fong, Jonathan Shade, Wayne Wooten, Brenden Schubert,
  Andrew Kensler, Stephen Friedman, Charlie Kilpatrick, Cliff Ramshaw, Marc
  Bannister, Brenton Rayner, Jonathan Brouillat, and Max Liani.
\newblock Renderman: An advanced path-tracing architecture for movie rendering.
\newblock \emph{ACM Trans. Graph.}, 37\penalty0 (3), August 2018.
\newblock ISSN 0730-0301.
\newblock \doi{10.1145/3182162}.
\newblock URL \url{https://doi.org/10.1145/3182162}.

\bibitem[M\"uller(2021)]{M_ller_2021_2}
Thomas M\"uller.
\newblock {tiny-cuda-nn}, 4 2021.
\newblock URL \url{https://github.com/NVlabs/tiny-cuda-nn}.

\bibitem[M\"{u}ller et~al.(2021)M\"{u}ller, Rousselle, Nov\'{a}k, and
  Keller]{M_ller_2021}
Thomas M\"{u}ller, Fabrice Rousselle, Jan Nov\'{a}k, and Alexander Keller.
\newblock Real-time neural radiance caching for path tracing.
\newblock \emph{ACM Trans. Graph.}, 40\penalty0 (4), jul 2021.
\newblock ISSN 0730-0301.
\newblock \doi{10.1145/3450626.3459812}.
\newblock URL \url{https://doi.org/10.1145/3450626.3459812}.

\bibitem[Yang et~al.(2023)Yang, Zhao, Luo, Wang, Sun, Li, Cai, and
  Jin]{yang2023}
Sipeng Yang, Yunlu Zhao, Yuzhe Luo, He~Wang, Hongyu Sun, Chen Li, Binghuang
  Cai, and Xiaogang Jin.
\newblock Mnss: Neural supersampling framework for real-time rendering on
  mobile devices.
\newblock \emph{IEEE Transactions on Visualization and Computer Graphics},
  pages 1--14, 2023.
\newblock \doi{10.1109/TVCG.2023.3259141}.

\bibitem[NVIDIA(2022)]{NvidiaDLSS3}
NVIDIA.
\newblock {NVIDIA} {Ada} {GPU} science, 2022.
\newblock URL
  \url{https://images.nvidia.com/aem-dam/Solutions/geforce/ada/ada-lovelace-architecture/nvidia-ada-gpu-science.pdf}.

\bibitem[M\"{u}ller et~al.(2022)M\"{u}ller, Evans, Schied, and
  Keller]{muller2022}
Thomas M\"{u}ller, Alex Evans, Christoph Schied, and Alexander Keller.
\newblock Instant neural graphics primitives with a multiresolution hash
  encoding.
\newblock \emph{ACM Trans. Graph.}, 41\penalty0 (4), jul 2022.
\newblock ISSN 0730-0301.
\newblock \doi{10.1145/3528223.3530127}.
\newblock URL \url{https://doi.org/10.1145/3528223.3530127}.

\bibitem[NVIDIA(2020)]{NvidiaAmpere}
NVIDIA.
\newblock {NVIDIA} {A100} tensor core {GPU} architecture, 2020.
\newblock URL
  \url{https://www.nvidia.com/content/dam/en-zz/Solutions/Data-Center/nvidia-ampere-architecture-whitepaper.pdf}.

\bibitem[Intel®(2021)]{OpenAIDenoise}
Intel®.
\newblock Intel® open image denoise, 2021.
\newblock URL \url{https://www.openimagedenoise.org/}.

\bibitem[Zhang et~al.(2021)Zhang, Manzi, Vogels, Dahlberg, Gross, and
  Papas]{zhang2021}
Xianyao Zhang, Marco Manzi, Thijs Vogels, Henrik Dahlberg, Markus Gross, and
  Marios Papas.
\newblock Deep compositional denoising for high-quality monte carlo rendering.
\newblock \emph{Computer Graphics Forum}, 40\penalty0 (4):\penalty0 1--13,
  2021.
\newblock \doi{https://doi.org/10.1111/cgf.14337}.
\newblock URL \url{https://onlinelibrary.wiley.com/doi/abs/10.1111/cgf.14337}.

\bibitem[Müller et~al.(2020)Müller, Rousselle, Novák, and
  Keller]{M_ller2020}
Thomas Müller, Fabrice Rousselle, Jan Novák, and Alexander Keller.
\newblock Neural control variates, 2020.
\newblock URL \url{https://arxiv.org/abs/2006.01524}.

\bibitem[Mildenhall et~al.(2021)Mildenhall, Srinivasan, Tancik, Barron,
  Ramamoorthi, and Ng]{Mildenhall2021}
Ben Mildenhall, Pratul~P. Srinivasan, Matthew Tancik, Jonathan~T. Barron, Ravi
  Ramamoorthi, and Ren Ng.
\newblock Nerf: Representing scenes as neural radiance fields for view
  synthesis.
\newblock \emph{Commun. ACM}, 65\penalty0 (1):\penalty0 99–106, dec 2021.
\newblock ISSN 0001-0782.
\newblock \doi{10.1145/3503250}.
\newblock URL \url{https://doi.org/10.1145/3503250}.

\bibitem[Zhi et~al.(2021)Zhi, Laidlow, Leutenegger, and
  Davison]{zhi2021inplace}
Shuaifeng Zhi, Tristan Laidlow, Stefan Leutenegger, and Andrew~J. Davison.
\newblock In-place scene labelling and understanding with implicit scene
  representation, 2021.

\bibitem[Blomqvist et~al.(2023)Blomqvist, Milano, Chung, Ott, and
  Siegwart]{blomqvist2023neural}
Kenneth Blomqvist, Francesco Milano, Jen~Jen Chung, Lionel Ott, and Roland
  Siegwart.
\newblock Neural implicit vision-language feature fields, 2023.

\bibitem[Poole et~al.(2022)Poole, Jain, Barron, and
  Mildenhall]{poole2022dreamfusion}
Ben Poole, Ajay Jain, Jonathan~T. Barron, and Ben Mildenhall.
\newblock Dreamfusion: Text-to-3d using 2d diffusion, 2022.

\bibitem[Rahaman et~al.(2018)Rahaman, Baratin, Arpit, Draxler, Lin, Hamprecht,
  Bengio, and Courville]{Rahaman2018}
Nasim Rahaman, Aristide Baratin, Devansh Arpit, Felix Draxler, Min Lin, Fred~A.
  Hamprecht, Yoshua Bengio, and Aaron Courville.
\newblock On the spectral bias of neural networks.
\newblock 2018.
\newblock \doi{10.48550/ARXIV.1806.08734}.
\newblock URL \url{https://arxiv.org/abs/1806.08734}.

\bibitem[Tancik et~al.(2020)Tancik, Srinivasan, Mildenhall, Fridovich{-}Keil,
  Raghavan, Singhal, Ramamoorthi, Barron, and Ng]{Tancik2020}
Matthew Tancik, Pratul~P. Srinivasan, Ben Mildenhall, Sara Fridovich{-}Keil,
  Nithin Raghavan, Utkarsh Singhal, Ravi Ramamoorthi, Jonathan~T. Barron, and
  Ren Ng.
\newblock Fourier features let networks learn high frequency functions in low
  dimensional domains.
\newblock \emph{CoRR}, abs/2006.10739, 2020.
\newblock URL \url{https://arxiv.org/abs/2006.10739}.

\bibitem[Sitzmann et~al.(2020)Sitzmann, Martel, Bergman, Lindell, and
  Wetzstein]{Vincent2020}
Vincent Sitzmann, Julien N.~P. Martel, Alexander~W. Bergman, David~B. Lindell,
  and Gordon Wetzstein.
\newblock Implicit neural representations with periodic activation functions.
\newblock \emph{CoRR}, abs/2006.09661, 2020.
\newblock URL \url{https://arxiv.org/abs/2006.09661}.

\bibitem[Bricman and Ionescu(2018)]{Bricman2018}
Paul~Andrei Bricman and Radu~Tudor Ionescu.
\newblock Coconet: A deep neural network for mapping pixel coordinates to color
  values, 2018.
\newblock URL \url{https://arxiv.org/abs/1805.11357}.

\bibitem[Minnen et~al.(2018)Minnen, Toderici, Covell, Chinen, Johnston, Shor,
  Hwang, Vincent, and Singh]{Minnen2018}
David Minnen, George Toderici, Michele Covell, Troy Chinen, Nick Johnston, Joel
  Shor, Sung~Jin Hwang, Damien Vincent, and Saurabh Singh.
\newblock Spatially adaptive image compression using a tiled deep network.
\newblock 2018.
\newblock \doi{10.48550/ARXIV.1802.02629}.
\newblock URL \url{https://arxiv.org/abs/1802.02629}.

\bibitem[Thies et~al.(2019)Thies, Zollhöfer, and Nießner]{Thies2019}
Justus Thies, Michael Zollhöfer, and Matthias Nießner.
\newblock Deferred neural rendering: Image synthesis using neural textures,
  2019.
\newblock URL \url{https://arxiv.org/abs/1904.12356}.

\bibitem[Kuznetsov et~al.(2021)Kuznetsov, Mullia, Xu, Hašan, and
  Ramamoorthi]{Kuznetsov2021}
Alexandr Kuznetsov, Krishna Mullia, Zexiang Xu, Miloš Hašan, and Ravi
  Ramamoorthi.
\newblock Neumip: Multi-resolution neural materials, 2021.
\newblock URL \url{https://arxiv.org/abs/2104.02789}.

\bibitem[Zeltner et~al.(2023)Zeltner, Rousselle, Weidlich, Clarberg, Novák,
  Bitterli, Evans, Davidovič, Kallweit, and Lefohn]{zeltner2023realtime}
Tizian Zeltner, Fabrice Rousselle, Andrea Weidlich, Petrik Clarberg, Jan
  Novák, Benedikt Bitterli, Alex Evans, Tomáš Davidovič, Simon Kallweit,
  and Aaron Lefohn.
\newblock Real-time neural appearance models, 2023.

\bibitem[Aila and Karras(2010)]{Aila2010}
Timo Aila and Tero Karras.
\newblock Architecture considerations for tracing incoherent rays.
\newblock In \emph{Proceedings of the Conference on High Performance Graphics},
  HPG '10, page 113–122, Goslar, DEU, 2010. Eurographics Association.

\bibitem[Pharr et~al.(2016)Pharr, Jakob, and Humphreys]{pharr2016physically}
Matt Pharr, Wenzel Jakob, and Greg Humphreys.
\newblock \emph{Physically based rendering: From theory to implementation}.
\newblock Morgan Kaufmann, 2016.

\bibitem[Wald et~al.(2014)Wald, Woop, Benthin, Johnson, and Ernst]{wald2014}
Ingo Wald, Sven Woop, Carsten Benthin, Gregory~S. Johnson, and Manfred Ernst.
\newblock Embree: A kernel framework for efficient cpu ray tracing.
\newblock \emph{ACM Trans. Graph.}, 33\penalty0 (4), jul 2014.
\newblock ISSN 0730-0301.
\newblock \doi{10.1145/2601097.2601199}.
\newblock URL \url{https://doi.org/10.1145/2601097.2601199}.

\bibitem[Benthin et~al.(2018)Benthin, Wald, Woop, and \'{A}fra]{Benthin2018}
Carsten Benthin, Ingo Wald, Sven Woop, and Attila~T. \'{A}fra.
\newblock Compressed-leaf bounding volume hierarchies.
\newblock In \emph{Proceedings of the Conference on High-Performance Graphics},
  HPG '18, New York, NY, USA, 2018. Association for Computing Machinery.
\newblock ISBN 9781450358965.
\newblock \doi{10.1145/3231578.3231581}.
\newblock URL \url{https://doi.org/10.1145/3231578.3231581}.

\bibitem[Keely(2014)]{Keely2014}
Sean Keely.
\newblock {Reduced Precision for Hardware Ray Tracing in GPUs}.
\newblock In Ingo Wald and Jonathan Ragan-Kelley, editors, \emph{Eurographics/
  ACM SIGGRAPH Symposium on High Performance Graphics}. The Eurographics
  Association, 2014.
\newblock ISBN 978-3-905674-60-6.
\newblock \doi{10.2312/hpg.20141091}.

\bibitem[LeGendre et~al.(2019)LeGendre, Ma, Fyffe, Flynn, Charbonnel, Busch,
  and Debevec]{LeGendre2019}
Chloe LeGendre, Wan-Chun Ma, Graham Fyffe, John Flynn, Laurent Charbonnel, Jay
  Busch, and Paul Debevec.
\newblock Deeplight: Learning illumination for unconstrained mobile mixed
  reality, 2019.
\newblock URL \url{https://arxiv.org/abs/1904.01175}.

\bibitem[Gupta et~al.(2015)Gupta, Agrawal, Gopalakrishnan, and
  Narayanan]{gupta2015deep}
Suyog Gupta, Ankur Agrawal, Kailash Gopalakrishnan, and Pritish Narayanan.
\newblock Deep learning with limited numerical precision, 2015.

\bibitem[Moshier(1992)]{moshier1992}
Stephen~L Moshier.
\newblock {Cephes mathematical library}, 1992.
\newblock URL \url{http://www.netlib.org/cephes/}.

\bibitem[Community(2018)]{blender2018}
Blender~Online Community.
\newblock \emph{Blender - a 3D modelling and rendering package}.
\newblock Blender Foundation, Stichting Blender Foundation, Amsterdam, 2018.
\newblock URL \url{http://www.blender.org}.

\bibitem[Mantiuk et~al.(2011)Mantiuk, Kim, Rempel, and Heidrich]{Mantiuk2011}
Rafa\l{} Mantiuk, Kil~Joong Kim, Allan~G. Rempel, and Wolfgang Heidrich.
\newblock Hdr-vdp-2: A calibrated visual metric for visibility and quality
  predictions in all luminance conditions.
\newblock \emph{ACM Trans. Graph.}, 30\penalty0 (4), jul 2011.
\newblock ISSN 0730-0301.
\newblock \doi{10.1145/2010324.1964935}.
\newblock URL \url{https://doi.org/10.1145/2010324.1964935}.

\bibitem[Fouladi et~al.(2022)Fouladi, Shacklett, Poms, Arora, Ozdemir,
  Raghavan, Hanrahan, Fatahalian, and Winstein]{Fouladi2022}
Sadjad Fouladi, Brennan Shacklett, Fait Poms, Arjun Arora, Alex Ozdemir, Deepti
  Raghavan, Pat Hanrahan, Kayvon Fatahalian, and Keith Winstein.
\newblock R2e2: Low-latency path tracing of terabyte-scale scenes using
  thousands of cloud cpus.
\newblock \emph{ACM Trans. Graph.}, 41\penalty0 (4), jul 2022.
\newblock ISSN 0730-0301.
\newblock \doi{10.1145/3528223.3530171}.
\newblock URL \url{https://doi.org/10.1145/3528223.3530171}.

\bibitem[Park et~al.(2019)Park, Florence, Straub, Newcombe, and
  Lovegrove]{park2019}
Jeong~Joon Park, Peter Florence, Julian Straub, Richard~A. Newcombe, and Steven
  Lovegrove.
\newblock Deepsdf: Learning continuous signed distance functions for shape
  representation.
\newblock \emph{CoRR}, abs/1901.05103, 2019.
\newblock URL \url{http://arxiv.org/abs/1901.05103}.

\bibitem[Graphcore(2023)]{ipuraylib2023}
Graphcore.
\newblock {IPU} ray library, 2023.
\newblock URL \url{https://github.com/graphcore-research/ipu-ray-lib}.

\bibitem[Valiant(1990)]{valiant1990bridging}
Leslie~G Valiant.
\newblock A bridging model for parallel computation.
\newblock \emph{Communications of the ACM}, 33\penalty0 (8):\penalty0 103--111,
  1990.

\bibitem[Cheatham et~al.(1996)Cheatham, Fahmy, Stefanescu, and
  Valiant]{cheatham1996bulk}
Thomas Cheatham, Amr Fahmy, Dan Stefanescu, and Leslie Valiant.
\newblock Bulk synchronous parallel computing—a paradigm for transportable
  software.
\newblock In \emph{Tools and Environments for Parallel and Distributed
  Systems}, pages 61--76. Springer, 1996.
\newblock URL \url{https://dash.harvard.edu/handle/1/26506457}.

\bibitem[Blackman and Vigna(2018)]{blackman2018scrambled}
David Blackman and Sebastiano Vigna.
\newblock Scrambled linear pseudorandom number generators.
\newblock \emph{arXiv preprint arXiv:1805.01407}, 2018.

\bibitem[Vigna(2019)]{vigna2019high}
Sebastiano Vigna.
\newblock It is high time we let go of the mersenne twister.
\newblock \emph{arXiv preprint arXiv:1910.06437}, 2019.

\bibitem[Graphcore(2021)]{GCDocs}
Graphcore.
\newblock Graphcore documents, 2021.
\newblock URL \url{https://docs.graphcore.ai/}.

\end{thebibliography}

\appendix

\section{Hardware Description}
\label{sec:hw}
Graphcore processors have not been adopted for rendering before so we present relevant details of the hardware and programming model. Whilst nominally described as an AI accelerator, IPUs are relatively general purpose parallel processors. Each IPU consists of a large number of homogeneous cores (called {\em tiles}) that synchronise in hardware using an embodiment of the bulk synchronous parallel (BSP) computation paradigm \cite{valiant1990bridging, cheatham1996bulk}. In this paradigm all cores compute, then synchronise at a barrier, then exchange data, and repeat. This computation model is well suited for expressing massively parallel computation without the programmer needing to micro-manage synchronisation and race conditions: essentially it is a hardware barrier synchronisation scheme. IPUs can be arranged into multi-IPU devices where the inter-IPU synchronisation is also achieved via a BSP sequence. Like GPUs, IPUs are accelerators, not standalone processors, and therefore require a host computer to run an operating system and orchestrate computation.

\subsection{IPU Processor Architecture}
\label{sec:IpuArch}

The GC200 processor contains 1472 tiles (cores) and each tile has a six barrel scheduled hardware threads called {\em workers}, each receiving one execution slot in round-robin fashion. This fixed schedule hides the SRAM latency, simplifies the compiler's view of each thread, and eliminates the need for out of order execution or branch prediction. The barrel contains an extra slot reserved for a supervisor thread which is responsible for initialisation of worker threads and inter-core communication. The workers can execute completely independent programs (each with their own control flow) and hence this processor is genuinely capable of multiple-instruction-multiple-data (MIMD) computations: a single GC200 can be running 8832 independent programs at once. Each tile also has the following features that are expanded upon below:
\begin{itemize}
	\item Private SRAM: 624KiB per tile, 897MiB aggregate per chip.
	\item A vector floating point unit (VFPU) capable of half and single precision calculations with (optional) stochastic rounding.
    \item An accumulating matrix product (AMP) unit (systolic array).
	\item A hardware random number generator.
	\item All to all data exchange with other tiles.
\end{itemize}

While the local SRAM could be viewed as a cache the BSP architecture and graph programming model (Section \ref{sec:ProgModel}) ensure its contents is completely deterministic. There is no memory hierarchy and tiles cannot issue an external memory fetches directly themselves. Instead, DRAM I/O requires a separate step in the BSP schedule. In effect, all tiles must sync before data can be streamed on/off chip. There is, however, an option to have a subset of tiles perform off-chip I/O while all the other tiles compute facilitating asynchronous host I/O or memory fetch pipelines.

The VFPU is 2-wide for float and 4-wide half precision FLOPs. The AMP unit achieves the device's peak floating point throughput and accelerates the large matrix multiplies and convolutions in deep learning.

The hardware random number generator (HW-RNG) is based on the xoshiro family \cite{blackman2018scrambled} and can directly produce uniformly distributed single precision floating point values with superior statistical properties to the popular Mersenne-Twister \cite{vigna2019high}.

Inter-tile communication is achieved by a proprietary graph compiler (Poplar) scheduling all of the program's communication in advance. This allows the inter-tile communication to be protocol free: no bandwidth is lost to addressing or packet headers. Inter-chip communication is packet based but still prearranged for efficiency and to maintain the BSP programming paradigm throughout the communication hierarchy.

\subsubsection{IPU System Architecture}

We will not go into full details of all possible IPU system configurations but we note that GC200s are usually available in groups of 4 in an IPU Pod 4: a 1U chassis which also contains the external DRAM (up to 450GB). These chips also have high-speed (320GB/sec) interconnects directly between themselves allowing for direct data exchange amongst their 3.5GiB of aggregate SRAM. Effectively, currently available IPU systems have traded external memory bandwidth for larger external memory size, targeting applications where larger internal SRAM, inter-core, and inter-chip communication bandwidth is more useful. It is important to take advantage of this trade-off when designing or selecting algorithms suitable for the current generation of IPU systems.

\subsection{IPU Programming Model}
\label{sec:ProgModel}

The core SDK for IPU programming consists of a C++ API for describing and compiling computational graphs (Poplar), a codelet (kernel) LLVM compiler (PopC), and a utility library that builds on these (Poplibs).

\subsubsection{The Compute Graph}
\label{sec:computegraph}

Poplar is used to construct a compute graph consisting of {\em vertices}, describing computations, connected by {\em edges}, describing data flow. Data can be n-way tensors of various floating point and integer data types. Arbitrary structures and arrays of plain old data can be transferred to the device in byte tensors which can then be reinterpret cast for use by the compute kernels. A computation graph is built by adding compute vertices to the program and connecting fields in those vertices to tensors to specify the data dependencies.

\subsubsection{Compute Kernels (Codelets)}
\label{sec:computekernels}
A compute kernel in Poplar is called a codelet. Codelets define the calculations performed by a worker at each compute vertex. Vertex code is written in C++ and compiled by PopC. These vertices can also be written using assembly for maximum performance, or optimised using inline assembly or intrinsics to access IPU hardware specific functionality.

\subsubsection{Programs and Execution}
\label{sec:PoplarPrograms}

Once the compute graph is constructed Poplar also needs a separate description of how to execute it in the form of the graph program. The simplest form of program is just to execute a group of vertices (compute set) but execution can also be arranged into sequences and loops allowing for complex and data dependent control flow within the fixed graph. This control flow is executed on device with no interaction from the host machine. The host tells the device which program to run then monitors when to sync with the device in the BSP schedule for I/O purposes. The final component needed for a complete program are data streams between the host and device or between the device and DRAM. Full details can be found in the Poplar SDK documentation \cite{GCDocs}.

\section{HDRI Dataset and Credits}
\label{sec:hdri-assets}

The HDRIs used to train the HDR-NIFs in this work were all distributed under the CC0 license:

\begin{itemize}
  \item \href{https://polyhaven.com/a/urban_alley_01}{urban\_alley\_01}, \href{https://polyhaven.com/a/borghese_gardens}{borghese\_gardens}, and \href{https://polyhaven.com/a/green_sanctuary}{green\_sanctuary} by Andreas Mischok
  \item \href{https://polyhaven.com/a/country_club}{country\_club} and \href{https://polyhaven.com/a/studio_small_09}{studio\_small\_09} by Sergej Majboroda
  \item \href{https://polyhaven.com/a/shanghai_bund}{shanghai\_bund} and \href{https://polyhaven.com/a/syferfontein_1d_clear}{syferfontein\_1d\_clear} by Greg Zaal 
\end{itemize}

\end{multicols}
\end{document}